\documentclass{SciPost}

\hypersetup{
    colorlinks,
    linkcolor={red!50!black},
    citecolor={blue!50!black},
    urlcolor={blue!80!black}
}

\usepackage[bitstream-charter]{mathdesign}
\DeclareSymbolFont{usualmathcal}{OMS}{cmsy}{m}{n}
\DeclareSymbolFontAlphabet{\mathcal}{usualmathcal}

\fancypagestyle{SPstyle}{
\fancyhf{}
\lhead{\colorbox{scipostblue}{\bf \color{white} ~SciPost Physics }}
\rhead{{\bf \color{scipostdeepblue} ~Submission }}

\fancyfoot[C]{\textbf{\thepage}}
}

\newcommand{\ket}[1]{|{#1}\rangle}

\newcommand{\beq}{\begin{equation}}
\newcommand{\eeq}{\end{equation}}

\usepackage{siunitx}
\usepackage{dcolumn}
\usepackage{orcidlink}
\usepackage{multirow}

\begin{document}

\pagestyle{SPstyle}

\begin{center}
{\Large \textbf{\color{scipostdeepblue}{
Collimation of dense atomic beams by Swept Velocity Shelving
}}}
\end{center}

\begin{center}
\textbf{
Francesca Famà\,\orcidlink{0009-0009-3162-8806}\textsuperscript{1,2,3$\star$},
Benedikt Heizenreder\,\orcidlink{0000-0001-5611-4144}\textsuperscript{2,3$\star$},
Ananya Sitaram\,\orcidlink{0000-0002-6548-4515}\textsuperscript{2,3},
Camila Beli\,\textsuperscript{2,3,4},
Robert J.C. Spreeuw\,\orcidlink{0000-0002-2631-5698} \textsuperscript{2,3},
Stefan A. Sch\"{a}ffer\,\orcidlink{0000-0002-5296-0332}\textsuperscript{2,3,4}
and Florian Schreck\,\orcidlink{0000-0001-8225-8803}\textsuperscript{2,3$\dagger$}
}
\end{center}

\begin{center}

{\bf 1} Istituto Nazionale di Ricerca Metrologica (INRiM), Turin, Italy \\
{\bf 2} Van der Waals-Zeeman Institute, Institute of Physics, University of Amsterdam, Amsterdam, The Netherlands \\
{\bf 3} QuSoft, Amsterdam, The Netherlands \\
{\bf 4} NNF Quantum Computing Programme, Niels Bohr Institute, University of Copenhagen, Copenhagen, Denmark

\vspace{0.5em}

$\dagger$ \href{mailto:SVS@strontiumbec.com}{\small SVS@strontiumbec.com}\\[4pt]
$\star$ \textit{These authors contributed equally to this work.}

\end{center}

\section*{\color{scipostdeepblue}{Abstract}}
\textbf{\boldmath{%
Engineering continuous, high-flux, and collimated atomic beams is a useful resource for metrology and material deposition. Developments in this area have been essential for the evolution of cold-atom based quantum experiments, yet the ubiquitous balanced-force methods such as transverse molasses cooling degrade at high atomic flux due to absorption-induced force imbalance. We introduce a collimation scheme that combines the use of a broadband transition for velocity shifting with a narrowband transition for velocity selection, enabling velocity-selective beam collimation without relying on balanced-power counter-propagating beams. Collimated atoms are shelved in a long-lived internal state, reducing the total light scattering and providing a state-heralded collimated beam. Simulations using strontium as a model system show a highly effective collimation process that does not suffer from absorption-induced force-imbalance and experimental results agree well with these predictions.}}

\vspace{\baselineskip}





\vspace{10pt}
\noindent\rule{\textwidth}{1pt}
\tableofcontents
\noindent\rule{\textwidth}{1pt}
\vspace{10pt}

\section{\label{sec:1}Introduction}
Beams of cold atoms are a highly valuable tool in the atomic physics community, where there has been a push towards continuous systems in recent years to decrease experimental cycle times and move beyond pulsed operation. Examples of this effort include continuous quantum simulators~\cite{chen_continuous_2022,Steady_state_magneto_optical_Escudero,boughdachi2025strontium}, atom interferometry~\cite{keit:91, kwo:20}, scalable quantum-computing architectures~\cite{Chiu2025,Atom_computing,Li2025_fast_continuous_coherent,Singh2022det}, and optical lattice clocks aiming to overcome the Dick effect through steady-state interrogation and approach 1/time scaling of clock uncertainty~\cite{katori_longitudinal_2021,Zero-Dead-Time_Operation_Biedermann,Zero-Dead-Time_Liu_2025,HeizenrederMOTSpec2025}. Similarly, cavity quantum-electrodynamics (QED) systems increasingly pursue continuous superradiant emission for applications in precision metrology and many-body physics~\cite{Meiser09,Liu:2020,schafer_continuous_2025,Modeling_swadheen,ElBadawi2025ContinuousSuperradiantLaser}. Finally, by achieving efficient and dense beams of collimated atoms, new applications within atom-surface interactions and nanofabrication could open up \cite{Gard13}.
\\
A common requirement shared by these approaches is the availability of atomic sources with high flux and narrow velocity distributions. In particular, active and continuous clocks critically rely on high-flux atomic beams with reduced Doppler broadening to enable long interaction times or stable cavity-atom interaction. Generating beams at high densities allows stronger atomic interaction or the realization of large-scale atomic systems \cite{Chiu2025,Gyger2024}, by increasing the atomic loading rate.
Comparable demands arise in precision molecular spectroscopy~\cite{vanHofslot2026BaFLaserCooling,Narrowline_Laser_Cooling_YO_Mehling_2025,Laser-cooled_BaF_Kogel_2025}, where narrow velocity distributions are essential to achieve high spectral resolution. The preparation of intense atomic beams that combine high flux with small velocity spread is therefore an enabling component across a range of fields.
\\
A standard approach to velocity narrowing is Doppler cooling, first demonstrated for atomic beam collimation in sodium~\cite{Phillips:85}. Optical molasses and its variants have since become a cornerstone of cold-atom physics. However, when applied to dense atomic beams, conventional molasses cooling encounters fundamental limitations. At high optical density, absorption leads to an intensity-imbalance between counter-propagating cooling beams, leading to a force imbalance. This hinders efficient cooling and results in broadened and shifted velocity distributions. Increasing the laser intensity can partially restore the force balance, but at the cost of power broadening, which directly increases the minimum achievable velocity spread~\cite{metcalf1999laser}. The additional scattering of photons from atoms at low velocities make these approaches costly from a photon efficiency point of view and materializes the standard Doppler cooling limit.
\\
More generally, Doppler cooling faces an intrinsic trade-off between capture velocity and final temperature. Broad optical transitions provide large capture ranges and strong scattering forces, while narrow transitions enable small velocity spreads, but drastically reduce capture efficiency~\cite{metcalf1999laser}. These competing requirements become particularly restrictive for hot, high-flux atomic beams, where both large capture velocity and narrow final velocity spread are simultaneously required. 
\\
Thus, such experiments require a fundamentally different strategy. Recent work has demonstrated that shielding atoms by driving on auxilliary transitions can regulate the effective scattering rate and mitigate losses~\cite{Höschele2023}, allowing enhanced spectroscopic performance~\cite{Li2025}. We introduce a beam-collimation approach that uses swept velocity shelving (SVS), a preparation scheme that allows for both dissipative laser-induced velocity control and atomic velocity selection while keeping both processes distinct. We do not rely on balanced light-forces but rather apply a one-sided force and conditional velocity-selection to achieve collimation of a heralded atomic ensemble. 
\\
The central idea is to employ a broad closed cycling transition for momentum control together with a narrow transition that incoherently transfers atoms within a selected velocity class into long-lived auxiliary states. Once shelved, these atoms no longer interact with the laser light. Velocity compression therefore does not rely on force balancing between counter-propagating beams but instead emerges from selective removal of atoms that satisfy a velocity class condition set by the experimentalist. In contrast to conventional optical molasses, where all atoms continuously participate in the light-atom interaction and collectively give rise to absorption-induced power imbalance, atoms fulfilling the velocity-selection criterion are rapidly removed from the optical cycle. As a result, reabsorption and power broadening of the cycling transition do not fundamentally limit the achievable velocity distribution.  
\\
In this work, we first analyze the limitations of conventional molasses cooling in dense atomic beams and introduce the theoretical framework of collimation by SVS. We then experimentally implement the scheme in a thermal strontium beam coupled to an optical cavity and quantitatively characterize the prepared atomic ensemble via normal-mode spectroscopy, allowing a direct comparison with optical molasses in a similar system~\cite{Fama24}.
\\
Our results demonstrate efficient narrowing of the transverse velocity distribution of a high-flux atomic beam in a regime where conventional Doppler cooling becomes absorption limited. This approach provides a route toward narrow-velocity, high-flux atomic sources, particularly relevant for continuous cavity-QED platforms and the realization of steady-state superradiant light sources.
\begin{figure}[t]
    \centering
    \includegraphics[width=\linewidth]{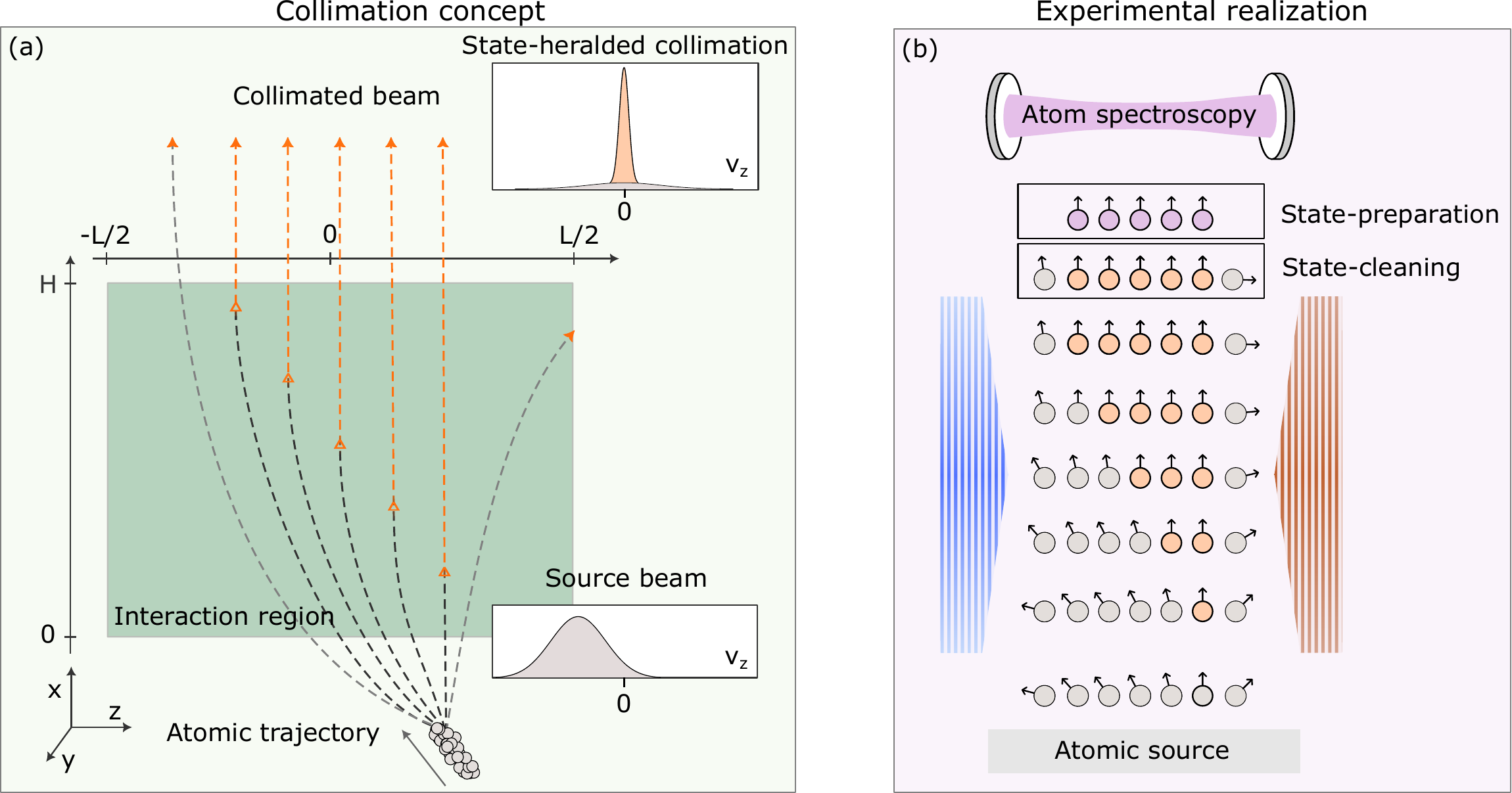}
    \caption{Minimal sketch of the collimation concept. (a) An atomic beam traverses a cooling region of dimensions $L \times H$, moving with vertical velocity $v_{x}$. The initial velocity distribution along $z$ is characterized by a mean $v_{z}^0$ and a broad width $\Delta v_{z}^0$, while the final distribution is characterized by $v_{z}^H$ and a narrow $\Delta v_{z}^H$. The source beam is assumed to have $v_{z}^0<0$ whereas the final distribution will be sharply peaked at $v_{z}^0=0$. Atoms within the final velocity distribution are heralded by their internal state, permitting selective addressing. (b) An experimental realization of the collimation concept with an atomic source, and pulsed beams of light addressing the atoms along the collimation direction. The heralded nature of the collimated atoms optionally enables a state-cleaning region, where uncollimated atoms are removed, and a state-preparation region, where the collimated atoms are transferred to a desired final internal state. We determine the normal-mode-splitting induced by the dense atomic beam in a linear optical cavity to characterize the beam temperature along the cavity direction and the atom number in the cavity mode.}\label{Fig:1}
\end{figure}

\section{Theory and Simulation}

We consider a system with atoms effusing from a source aperture to some region of experimental interest over a distance $H$, as illustrated in Fig. \ref{Fig:1}. Within this region, we consider the atomic beam to be traveling in the $x$-direction with a mean forward velocity much larger than the velocity spread along $y$ and the collimation direction $z$. For simplicity, we don't model the motion along $y$.

The $i$'th atom is characterized by its velocities $v_{x}^i$, which we assume to be constant, and $v_{z}^i(x)$. The constant $v_{x}^i$ fixes the atomic interaction time across the height $H$. We explore the final velocity-distribution $v_{z}(H)$ after traversing the collimation section as a function of the uniform atomic density $n$.

\subsection{Optical Molasses} 
The most well-known and widely applied technique for atomic cooling is optical molasses \cite{Will98, Lett89, Li2018}. It consists of employing power-balanced counter-propagating red-detuned laser beams, characterized by a detuning $\delta$ and wavevector $k$, to produce a viscous, velocity-dependent radiation pressure force that slows and cools atoms. In this context, we note two fundamental limitations that restrict the performance of a conventional molasses in high-flux regimes. First, a photon budget bottleneck arises because the Doppler cooling force remains active at zero velocity; atoms already brought to rest continue to scatter photons at a resonant rate throughout their transit, wasting a large fraction of the available laser power.
Second, significant absorption of the cooling beams leads to an imbalance in optical forces, which reduces the overall cooling efficiency and results in an increased final value of $\Delta v_{z}^H$ \cite{DAL88}.
One strategy to compensate for this absorption is to increase the intensity of the cooling beams. This approach introduces power broadening of the atomic transition, $\Gamma_\mathrm{eff} = \Gamma\sqrt{1 + s}$, with $s$ being the saturation parameter. This raises the minimum achievable Doppler temperature (or velocity $v_\text{sw}=\Gamma_\mathrm{eff}/k$). As a consequence, the final $\Delta v_{z}^H$ becomes higher than what could be achieved using the low-saturation regime with low optical density, ultimately limiting the cooling efficiency.

\begin{figure}[tbp] 
\centering 
\includegraphics[width=1\linewidth]{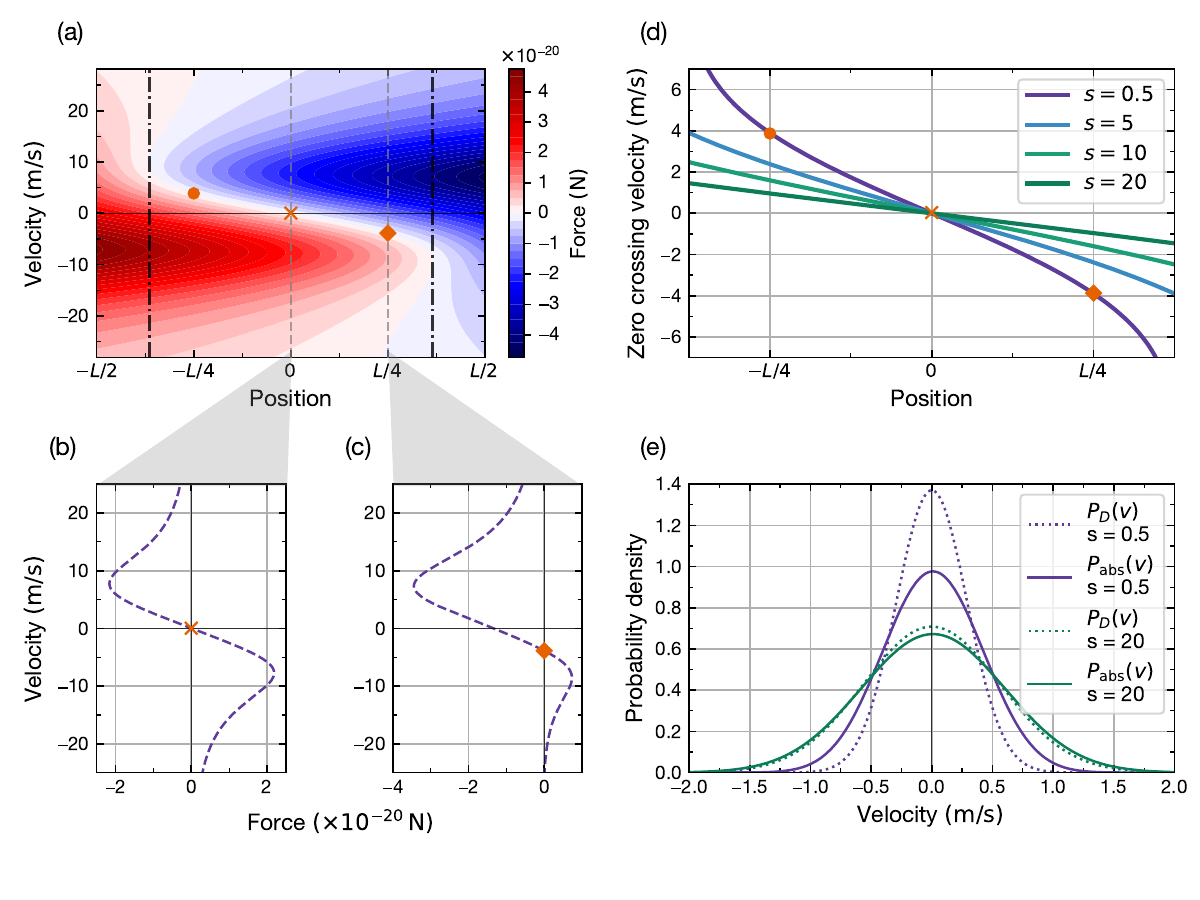} 
\caption{One–dimensional optical molasses in the presence of absorption by a dense atomic beam.
    \\
    (a) Cooling force $F(v,z)$ as a function of atomic velocity and transverse position $z$ along the molasses axis for an unsaturated configuration ($s=0.5$), calculated for the ${}^1S_0 \rightarrow {}^1P_1$ transition in Sr at detuning $\Delta=-0.5\,\Gamma$. The atomic density is $n=\qty{2e11}{cm^{-3}}$, and $L = \qty{1}{mm}$. Absorption induces a spatial imbalance between the counter-propagating beams and modifies the velocity-dependent cooling force. The orange symbols (disk, cross, and diamond) mark the velocity towards which atoms at the given position are cooled.
    \\
    Callouts show force profiles at these selected positions:
    (b) beam center ($z=0$), where balanced intensities result in zero-force at $v=0$ (orange cross) and symmetric capture velocities;
    (c) displaced position ($z=L/4$), where absorption produces an intensity imbalance that shifts the zero-force velocity to a finite velocity (orange diamond) and generates a strongly asymmetric capture range. The symmetric position ($z=-L/4$) exhibits the opposite velocity shift (orange disk).
    This spatial dependency of the cooling centers results in a broader velocity distribution after the molasses phase.
    \\
    (d) Spatial evolution of the stable velocity (zero-force point satisfying $\partial F/\partial v<0$) for different saturation parameters. Increasing saturation suppresses absorption-induced imbalance and compresses the spatial variation of the equilibrium velocity.
    \\
    (e) One-dimensional steady-state velocity distributions. Dotted curves show Doppler-limited Gaussian distributions expected for ideal balanced molasses at $s=0.5$ and $s=20$. Solid curves include absorption-induced imbalance, resulting in broadened velocity spread and increased effective temperatures. At large saturation, power broadening dominates the velocity spread.
    }
\label{Fig:2} 
\end{figure}

To illustrate this limitation, we evaluate the effective force of an unbalanced optical molasses at various axial positions $z$ within the cooling region of length $L$ (defined from $z = -L/2$ to $z = L/2$).
We model the local absorption of the counter-propagating cooling beams by assuming a uniform atomic density of $n=\qty{2e11}{\mathrm{cm^{-3}}}$ and an $x$-independent absorption profile. 
We assume that both laser beams are linearly polarized and only drive pi transitions \cite{Weiss:89}.

The bulk absorption coefficient for a specific velocity class $v_z$ is given by $\alpha(v_z) = n(v_z) \sigma(v_z)$, where the scattering cross-section $\sigma(v_z)$ is defined as:
\begin{align}
    \sigma &= \sigma_0 \frac{1}{1 + 4(\Delta(v_z)/\Gamma)^2 + s(z)},  \label{eq:cross_sec} \\
    \sigma_0 &= \frac{3 \lambda^2}{2 \pi}.
\end{align}
Here $\Delta(v_z) = \delta-k v_z$ is the Doppler shifted velocity-dependent detuning, and $s(z)$ is the position dependent saturation parameter. To account for the entire ensemble, the effective absorption coefficient $\bar{\alpha}$ is obtained by averaging $\alpha(v_z)$ over the atomic velocity distribution $f_v(v_z)$:
\begin{equation}
\bar{\alpha} = n \int_{-\infty}^{\infty} f_v(v_z) \sigma(v_z) \, \mathrm{d}v_z
\end{equation}

As the cooling beams traverse the medium along the $z$-axis (defined from $z = -L/2$ to $z = L/2$), they undergo Beer-Lambert attenuation governed by this ensemble-averaged $\bar{\alpha}$. Assuming symmetric absorption profiles for the counter-propagating pairs, we model the local saturation parameters for the right-propagating ($s_+$) and left-propagating ($s_-$) beams as:
\begin{align}s_+(z) &= s_0 e^{-\bar{\alpha} (z + L/2)}, \\
s_-(z) &= s_0 e^{-\bar{\alpha} (L/2 - z)},\end{align}
where $s_0$ is the initial unattenuated saturation parameter at the edges of the cloud. These position-dependent values, break the intensity symmetry of the molasses. 

Finally, $s_+(z)$ and $s_-(z)$ are fed into a semiclassical simulation using the PyLCP package \cite{Eck:22}, which employs rate equations to track photon scattering rates and compute the net mechanical force. We evaluate the cooling force immediately downstream from the source at $x=0$. For this calculation, we consider a strontium oven operating at approximately $500^\circ\text{C}$ equipped with a typical microcapillary nozzle array. The resulting geometric spatial filtering restricts the initial velocity distribution along the $z$-axis to a narrowed Gaussian profile with a standard deviation of $\sigma_v = \qty{28}{m/s}$. This width intentionally spans roughly twice the characteristic capture velocity of a standard \qty{461}{nm} optical molasses ($\approx \Gamma/k \approx \qty{14}{m/s}$), providing a good regime to investigate how laser attenuation impacts the cooling. For an initial $s_0 = 0.5$, this velocity distribution leads to a severe beam absorption of approximately $90\%$ across a cooling region of length $L = \qty{1}{mm}$.

The simulated net force is presented in Fig.~\ref{Fig:2}~(a). 
When moving away from the center of the atomic region, the velocity at which the force is zero shifts away from $v_z=0$ due to an imbalance between the cooling beam intensities (Fig.~\ref{Fig:2}~(b-c)). For a red-detuned optical molasses, these zero-force points are stable equilibria, as the stability condition $\frac{\partial \text{Force}}{\partial v_z} < 0$ is satisfied. Consequently, the atoms cool to a shifted local equilibrium velocity $v_0(z)$ (Fig.~\ref{Fig:2}~(d)).
Beyond a certain distance, no zero-crossing exists in the force profile, and all velocities will experience a force towards the center of the atomic region. Note that this can lead to advantages in some systems that require beam compression rather than atomic collimation and cooling.

In Figure~\ref{Fig:2}~(e) we show the Doppler distribution expected for an ideal balanced molasses (purple dotted curve) compared to an approximation of the distribution after the cooling stage when accounting for the force imbalance (solid purple curve). Rather than providing a rigorous full simulation, this calculation is intended for illustrative purposes to qualitatively demonstrate the limits of this molasses technique.
The solid purple curve is obtained by integrating the local Doppler profiles over the spatial distribution of $v_0(z)$, assuming that the cooling forces at $x=0$ hold for all $x$ and that the extent of the molasses is sufficient to reach equilibrium. This comparison shows an inhomogeneous broadening caused by the spatially dependent velocity shifts. Here we approximate the local Doppler distributions as having all the same width; in practice, the beam imbalance also reduces the local cooling efficiency by reducing the force gradient $dF/dv$, which leads to further broadening of the total distribution.
Furthermore, this model assumes an $x$-independent absorption profile. For a full description, one must account for the progressive evolution of the velocity distribution during cooling and the spatial trajectories of the atoms as they propagate through the laser fields. For example, while compression along the $z$-cooling axis increases the local density, any spatial expansion along $z$ preserves the column density encountered by the cooling beams, and thus leaves the total absorption unchanged. In contrast, spatial expansion along the uncooled $y$-axis allows atoms to escape the volume of the $z$-cooling beams, reducing the effective density $n$ and decreasing absorption further downstream. If two-dimensional transverse cooling ($y$ and $z$) is employed, this density decay is removed.

The spatial dependence of the zero-force point is plotted for various saturation parameters in Fig.~\ref{Fig:2}~(d). Although the slope is evident at lower intensities, it can be mitigated by using a higher laser intensity (increasing the saturation parameter). From the absorption we see that in the limit where $s \gg 1$ the scattering cross section approaches $\sigma_0/s$ and the medium becomes more transparent. The resulting power broadening raises the Doppler-limited temperature ($T \propto \Gamma_\text{eff}$ \cite{foot05}) resulting in a wider final velocity distribution. The high-saturation case is shown in Fig.~\ref{Fig:2}~(e) where the final probability density (solid green curve) approaches the expected Doppler-limit (dotted green curve) for a saturation broadened system. The Doppler cooling limit of optical molasses is also representative of the unwanted photon scattering of slow atoms. This addional absorption and rescattering process increasese the experimental power requirements and the intensity of unwanted secondary scattering.

Ultimately, this analysis demonstrates that attempting to overcome absorption by simply increasing the laser intensity is inherently self-defeating; while it forces the medium to become transparent, it simultaneously limits the cooling efficiency through power broadening. Furthermore, this elevated intensity compounds the photon budget bottleneck by forcing the already-cooled atoms to scatter redundant photons at a higher rate.

\subsection{Swept velocity shelving}
\label{sec:velocity-swept electron shelving}
\subsubsection{SVS operating principle \label{subsec:SVSOperatingPrinciple}}

\begin{figure}[t]
    \centering
    \includegraphics[width=1.\linewidth]{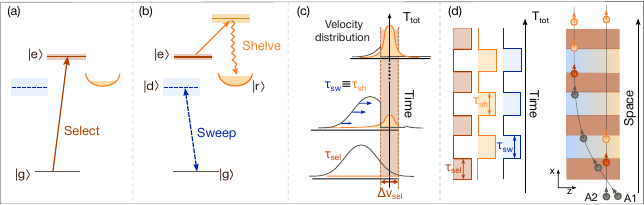}
    \caption{Schematic of the swept velocity shelving (SVS) technique. (a–b) Atomic transitions driven by the applied laser fields. Atoms start out in the ground state $\ket{g}$ and are transferred to $\ket{e}$ if they fall within the selected velocity range by a pulse of the selection beam. By subsequently driving $\ket{g} \rightarrow \ket{d}$ by a pulse of the sweeping beam, new atoms can enter into this range. Atoms in $\ket{e}$ are incoherently transferred to the reservoir state $\ket{r}$ via laser driving and spontaneous decay. (c) Schematic of time evolution of targeted atomic velocity distributions in $\ket{g}$ (gray) and $\ket{r}$ (orange). The atomic distribution is swept through the selected velocity window and atoms are transferred to the reservoir state when they fulfill the velocity selection criterium. (d) Laser pulse sequence experienced by individual atoms. (Left) Time sequence of the selection ($\tau_\text{sel}$), sweeping ($\tau_\text{sw}$), and shelving ($\tau_\text{sh}$) beam pulses. (Right) Spatial representation of the sequence, illustrating the trajectories and state-shelving behavior for a slow (A2) and a fast (A1) atom.
    For strontium the states correspond to: $\ket{g} \rightarrow {}^1S_0$, $\ket{d} \rightarrow {}^1P_1$, $\ket{e} \rightarrow {}^3P_1$ and $\ket{r} \rightarrow {}^3P_0$ and ${}^3P_2$. The atoms decay into the reservoir states via spontaneous emission ${}^3S_1$.
    \label{Fig:3}}
\end{figure}

In this work, we introduce an alternative method for preparing a cold sample of atoms starting from a broad thermal distribution.
The method relies on a controlled sequence of operations between three laser beams that address three distinct atomic transitions. Using these lasers, we can program the mean velocity $v_0$ and width $\Delta v$ of the final atomic distribution. A highly saturated broad transition is used to exert a force on the atoms in one direction (the "sweep" beam), a narrow transition is used to selectively address atoms within a chosen velocity class (the "selection" beam), and a third transition is used to transfer the selected atoms into a long-lived, non-interacting state, effectively serving as a reservoir (the "shelving" beam); see Fig.~\ref{Fig:3}~(a, b). 

These three beams are switched on and off in an optimized time-sequence. First, a pulse of the selection beam excites atoms around $v_0$ to the state $\ket{e}$. Immediately afterward, the shelving and sweeping beams are applied simultaneously: the shelving beam optically pumps $\ket{e}$ atoms to an intermediate state from which they spontaneously decay into the long-lived reservoir state $\ket{r}$, while the sweeping beam addresses the atoms in $\ket{g}$ to sweep their velocity distribution to overlap with the $v_0$ bin. Repeating this cycle allows for the stepwise accumulation of atoms that meet the condition $v_z\leq\Delta v_\text{sel}$ in the reservoir state $\ket{r}$, see Fig.~\ref{Fig:3}~(c). 

A key advantage of this one-sided force geometry is that, unlike in an optical molasses, beam absorption does not broaden the final velocity distribution, allowing the final velocity spread $\Delta v$ to reach the selection range $\Delta v_\text{sel}$.
Furthermore, tuning the selection laser's detuning allows us to center the distribution at any arbitrary mean velocity $v_0$. 

The velocity selection range $\Delta v_\text{sel}$ is primarily determined by the natural decay rate, the selection laser intensity, and the pulse duration. The latter two govern the power broadening (characterized by the saturation parameter $s_{\text{sel}}$) and Fourier broadening (governed by the inverse of the pulse duration $1/\tau_{\text{sel}}$). While the natural linewidth and power broadening result in a Lorentzian profile, the square temporal shape of the selection pulse introduces a $\text{sinc}^2$ line shape. As the pulse-area approaches an ideal $\pi$-pulse, the distribution becomes Lorentzian-like and narrow. \\
To capture the combined effect of these broadenings, the actual final velocity spread $\Delta v_z$ is parameterized by an effective Lorentzian width $\gamma_{\text{eff}} \sim k_{\text{sel}} \Delta v_z$, where $k_{\text{sel}}$ is the wavenumber of the selection transition.

Because power and Fourier broadening depend directly on intensity and duration, operating near a $\pi$-pulse offers programmable control over the selection bandwidth:
\begin{equation}
k_{\text{sel}} \Delta v_\text{sel} \sim \gamma_\mathrm{sel}\sqrt{1 + s_\mathrm{sel}} + \frac{1}{\tau_\mathrm{sel}}, 
\label{Eq:lin_broad}
\end{equation}
where $\gamma_\text{sel}$ is the natural linewidth of the ground-to-excited state pumping transition $\ket{g} \rightarrow \ket{e}$. 
Since the subsequent shelving step is fast and broad, the overall transfer efficiency of the sequence is governed primarily by the selection stage. Maximizing this transfer efficiency, achieved near an optimal $\pi$-pulse condition, ensures that the effective velocity width directly matches the ideal selection bandwidth ($\Delta v_\text{sel} \approx \Delta v_z$).

In particular , for long pulse durations, the power-broadened contribution $\gamma_\text{sel}\sqrt{1+s_\text{sel}}$ dominates, enabling straightforward tuning of the velocity spread $\Delta v_z$ simply by varying the laser power. Conversely, the Fourier term ($1/\tau_\text{sel}$) accounts for the finite interaction time $\tau_\text{sel}$ that each atom spends in the selection stage. This duration is governed by the vertical atomic velocity $v_x$ alongside the sequence switching frequency $f$, duty cycle $d$, and laser beam height $H$ (assumed equal to the interaction region height; see Fig.~\ref{Fig:1}(a)). Denoting the individual stage durations as $\tau_\text{sel}$ (selection), $\tau_\text{sw}$ (sweeping), and $\tau_\text{sh}$ (shelving, with $\tau_\text{sh} \equiv \tau_\text{sw}$; see Fig.~\ref{Fig:3}(b-c)), it follows that $f = 1/(\tau_\text{sel} + \tau_\text{sw})$ and $d = \tau_\text{sel}f$. Thus, a duty cycle of $70\%$ corresponds to a selection stage longer than the sweeping stage. Finally, the total sequence duration is given by $T_\text{tot} = N(\tau_\text{sw} + \tau_\text{sel}) = N/f$, where $N$ is the number of SVS periods. This parameterization directly connects the experimental control settings to the sequence dynamics.

As noted above, the final velocity spread is negligibly affected by the shelving transition. This transition is chosen to be sufficiently broad to rapidly transfer atoms to an intermediate state, from which they spontaneously decay into the reservoir state. Crucially, this spontaneous decay introduces the dissipation needed for phase-space cooling. As scattered photons carry entropy away into the radiation field, the process acts as a "one-way wall"~\cite{Rai08, Met08}, irreversibly trapping atoms in a narrow velocity spread without re-heating those already selected.

Figure~\ref{Fig:3}(d) illustrates the sequence experienced by each atom as it travels through the SVS region. Fast atoms (A1 in figure) undergo multiple sweeping stages until their velocity satisfies $|v_\text{Ai}| < \Delta v_\text{sel}/2$. If the required number of sweeping stages is less than $N-1$, the atom (e.g. A2 in Fig.~\ref{Fig:3}(d)) will be pumped by a selection stage and transferred to the reservoir. Thus, the number of sweeping stages $N$, together with the maximum deceleration per stage $-a_z^\text{max}$, determines the maximum initial transverse velocity that is effectively collimated and transferred. This scheme is inherently photon-efficient: unlike an optical molasses, where cooled atoms continuously scatter photons and suffer from recoil heating, atoms here scatter only the minimum number of photons required to reach the target velocity before being shelved.

\subsubsection{SVS numerical simulation\label{subsec:SVSNumericalSimulation}}

We simulate our SVS technique for the special case of strontium atoms to obtain real-world performance estimations for the collimation and transfer efficiency, which we compare to the experiment in Sec.~\ref{sec:ExpResults}. We begin by evaluating the transfer efficiency to the reservoir state, $\eta$, defined as the fraction of atoms within the targeted velocity class $\Delta v_\text{sel}$ that are successfully transferred, as a function of the sequence frequency $f$ and duty cycle $d$. For the optimal parameter pair $(f, d)$, we then evaluate the resulting velocity spread $\Delta v_z$ and compare it directly to the predicted selection range $\Delta v_\text{sel}$. Finally, we test the scheme at two different atomic densities to validate its performance under varying conditions.

We use the broad blue transition ${}^1S_0 \rightarrow {}^1P_1$ with linewidth $2\pi \times$\qty{30.5}{MHz} for sweeping, the narrow red transition ${}^1S_0 \rightarrow {}^3P_1$ with linewidth $2\pi \times$\qty{7.5}{kHz} for selecting, and the long-lived states ${}^3P_{2/0}$ as reservoir states for shelving. The population is transferred from ${}^3P_1$ to the reservoir states using the ${}^3P_1 \rightarrow {}^3S_1$ transition with natural linewidth $2\pi \times$\qty{4.3}{MHz}, addressed by the shelving laser. Once the atoms are in ${}^3S_1$, they spontaneously decay via the ${}^3S_1 \rightarrow {}^3P_{2/0}$ transitions, with rates $2\pi \times$\qty{6.68}{MHz} and $2\pi \times$\qty{1.42}{MHz}, respectively.

We use rate equations to simulate the sweeping stage, whereas for the selection and reservoir stages, Bloch equations are employed.
To model the output of an angled microtube array ($l = \qty{8}{mm}$, $r = \qty{150}{\micro m}$) operating at $T = 500^\circ\text{C}$, $N_\text{sim} = 50$ atoms are initialized in the ${}^1S_0$ state. Their initial positions $z$ are sampled uniformly over $[-L/2, L/2]$. Initial velocities $v_z$ are drawn from a Gaussian distribution centered at $v_z^0 = \qty{5}{m/s}$, with a width set by the microtube collimation geometry and truncated to $v_z \in [-5, v_\text{max}]\text{ m/s}$. This range reflects typical experimental conditions: atoms with negative velocities are accelerated away from the selection window, while the upper bound $v_\text{max}$ represents the maximal velocity that can be successfully decelerated into the selection range for a finite interaction height $H = \qty{2.5}{cm}$ and saturation parameter $s_\text{sw}$. 

As described above, the sequence consists of three operations. First, atoms within a velocity bandwidth $k_{\text{sel}}\Delta v_\text{sel}$ around a target velocity $v_z^\text{sel}$ are selected by excitation to ${}^3P_1$, with the bandwidth determined by the laser detuning and saturation parameter. In our simulations, we set $v_z^\text{sel} = 0$ (resonant light) and $s_\text{sel} = 10^4$, yielding a power-broadened Doppler width of $2\pi \times \qty{750}{kHz}$ (the first term in Eq.~\ref{Eq:lin_broad}). Second, this selected population is transferred to the intermediate state ${}^3S_1$ using $s_\text{sh} = 4.4$. From ${}^3S_1$, atoms spontaneously decay into ${}^3P_2$, ${}^3P_0$, or ${}^3P_{1, m_j = \pm1}$. While atoms falling into ${}^3P_2$ and ${}^3P_0$ are safely shelved in the reservoir, those decaying back into ${}^3P_{1, m_j = \pm1}$ remain active within the four-level optical Bloch equation system.\\
To reduce computational load when modeling this transfer, we adopt a fixed-efficiency approximation for the decay into the reservoir. This simplification is justified because the shelving transition linewidth is significantly broader than the selected velocity distribution, making the transfer efficiency effectively uniform across the ensemble. We validate this model against full numerical simulations using the complete 13-level optical Bloch equations for both pumping and shelving. Over representative sampled values of $f$ and $d$ (simulated for $N_\text{at} = 50$ atoms), the fixed-efficiency model shows agreement with the full 13-level calculation. 
\begin{figure}
    \centering
    \includegraphics[width=1.\linewidth]{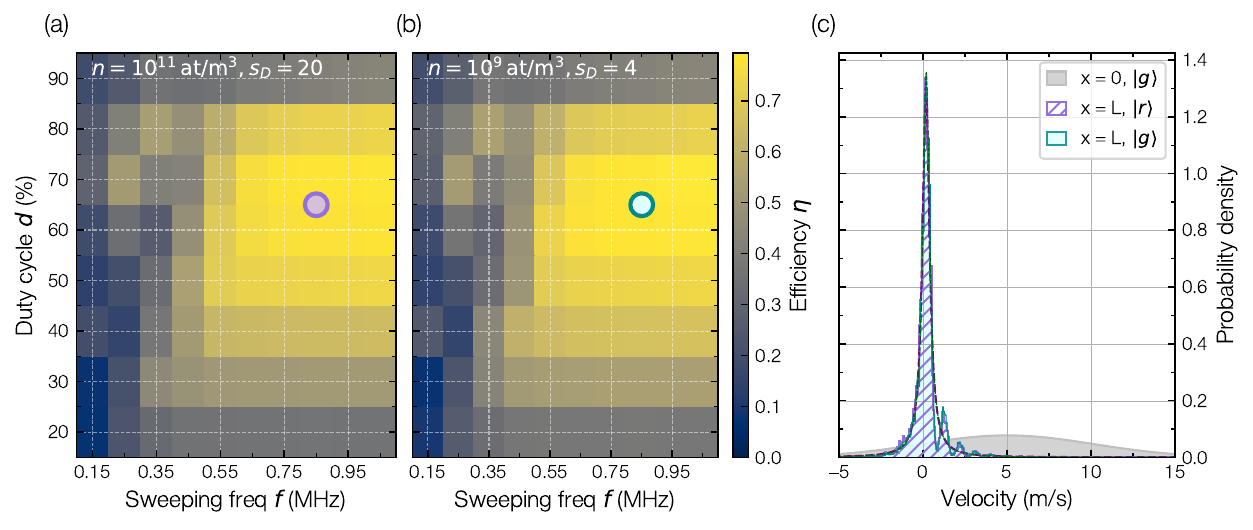}
    \caption{SVS efficiency simulations $\eta$ for (a) $n = \qty{2e11}{at/m^3}$, $s_\text{sw} = 20$ and (b) $n = \qty{2e9}{at/m^3}$, $s_\text{sw} = 4$ for $N = 42$ atoms as a function of sweeping frequency $f$ and duty cycle $d$. The circular marker indicates the high-efficiency operating point ($f = \qty{850}{kHz}$, $d = 50\%$). (c) Velocity probability density distributions. The assumed initial ground-state distribution is shown at the entry point ($x = 0$), while the final distribution of the shelved state, simulated with the optimized switching parameters $f = \qty{859}{kHz},\, d = 0.85\%$ using $N = 600$ atoms, is recorded at $x = L$. Dashed lines represent Lorentzian fits to the numerical data. The fitted velocity spread is 0.48(1) m/s for the purple high-density distribution (hatched) and 0.50(1) m/s for the green low-density distribution (filled).}
    \label{Fig:4trial}
\end{figure}

Finally, during the sweeping stage, the sweeping beam is applied to the remaining ground-state (${}^1S_0$) atoms with a saturation parameter $s_\text{sw}$. These remaining ground-state atoms, along with the unshelved population returning from ${}^3P_{1, m_j = \pm1}$, form the updated ensemble for the subsequent selection and shelving cycle. 

To determine the optimal sweeping parameters $(f, d)$, we start by evaluating the total transfer efficiency $\eta$, defined as the fraction of atoms accumulated in the shelving states. 
We perform this analysis for two atomic densities: $n = 2 \times 10^{11} \, \text{at/m}^3$ and $n = 2 \times 10^9 \, \text{at/m}^3$. The higher density corresponds to the flux directly at the nozzle exit ($x = -20\text{ mm}$), where we have shown the performance limits of conventional optical molasses. The lower density better represents the conditions in our upcoming experimental realization, accounting for the spatial expansion of the atomic beam along the $y$ direction during its drift from the nozzle exit (at $x=-20$\,mm) to the start of the collimation region ($x = \qty{0}{mm}$). (This travel is required in our setup to accommodate experimental components such as heat shields).
For the lower-density case, we set the sweeping beam saturation parameter to $s_\text{sw} = 4$, chosen specifically to match our experimental implementation.
This choice yields a maximum decelerated velocity of $v_\text{max} \approx \qty{32}{m/s}$, establishing an initial velocity range of $v_z \in [-5, 32]$\,m/s. To enable a direct comparison between the two density regimes, this initial velocity range is held fixed for the high-density simulation ($n = 2 \times 10^{11} \, \text{at/m}^3$), while the sweeping parameter is scaled upward to $s_\text{sw} = 20$ to compensate for the increased beam attenuation across the cloud.

The results are displayed in Fig.~\ref{Fig:4trial}(a) for an atomic density $n = \qty{2e11}{at/m^3}$ and Fig.~\ref{Fig:4trial}(b) for $n = \qty{2e9}{at/m^3}$. \\
We first note that both show a transfer efficiency exceeding $50\%$. For an untilted (symmetric) velocity distribution, the theoretical upper bound is strictly $50\%$, because atoms with negative velocities accelerate away from the selection window and are selected very inefficiently. However, the presence of an initial velocity tilt, combined with an optimized beam orientation, enables higher total efficiency.

To understand the location of the optimal $(f, d)$ parameters, we first consider the unperturbed, single-atom limit.
If the selection stage would be applied in isolation, the optimal parameter combinations $(f, d)$ would be those satisfying the $\pi$-pulse condition for a pulse duration $\tau_{\text{sel}} = d/f$:
$$\frac{d}{f} \gamma \sqrt{\frac{s_\text{sel}}{2}} \approx \pi.$$
Conversely, the sweeping stage independently achieves optimal transfer when each sweep decelerates atoms across the full selected velocity range $\Delta v_\text{sel}$ during the duration $\tau_{\text{sw}} = (1-d)/f$:
$$\Delta v_\text{sel} \sim \frac{\hbar k_\text{sw} \gamma_\text{sw}}{2m}\frac{s_\text{sw}}{ 1 + s_\text{sw}}\tau_\text{sw},$$
with the wavenumber of the sweeping transition $k_\text{sw}$, its linewidth $\gamma_\text{sw}$ and mass $m$ of $^{88}\text{Sr}$. \\
For our experimental parameters, these independent conditions suggest an optimum near $(f, d) \sim (\qty{150}{kHz}, 87\%)$. 
In practice, however, the overall efficiency is governed by the coupled interplay between selection, sweeping, and optical depth (atomic absorption). Because the sweeping force depends strongly on the local laser intensity, beam attenuation across the dense atomic ensemble reduces the effective saturation parameter inside the cloud. To compensate for this reduced force and still sweep atoms across $\Delta v_\text{sel}$, a longer sweeping duration $\tau_\text{sw}$ is required. As seen in the simulation map in Fig.~\ref{Fig:4trial}(a), this shifts the optimal duty cycle $d$ to lower values (around $60$ to $70\%$) than predicted by the single-atom model. Furthermore, because the ensemble transfer relies on a multi-pass process rather than a single ideal sweep, higher repetition frequencies $f$ are required to maintain steady-state transfer efficiency as atoms transit through the interaction region. 

This higher operating frequency implies that the final velocity spread is influenced by Fourier broadening from the short selection pulses. To verify that the optimized parameters yield a final velocity spread compatible with the selection range, we simulate the output velocity probability density distribution for $N_\text{at} = 600$ atoms at the optimal sweeping parameters $(f, d) = (\qty{850}{kHz}, 65\%)$ (indicated by the purple/green circular marker in Fig.~\ref{Fig:4trial}(a)/(b)). From Eq.~\eqref{Eq:lin_broad}, the expected selection bandwidth under these conditions is $\Delta v_\text{sel} \sim \qty{0.66}{m/s}$.
The resulting distribution from the simulation is shown as the purple/green curve in Fig.~\ref{Fig:4trial}(c). As anticipated from the Fourier transform of a square selection pulse, the velocity profile exhibits a characteristic $\operatorname{sinc}^2$ shape. Nevertheless, power broadening smooths the lineshape such that a Lorentzian distribution provides a good approximation, yielding a fitted velocity spread of $\Delta v_z = \qty{0.50(1)}{m/s}$ ($\gamma_\text{eff} = 2\pi \times \qty{725(15)}{kHz}$) for the high-density simulation (hatched purple area), and $\Delta v_z = \qty{0.48(1)}{m/s}$ ($\gamma_\text{eff} = 2\pi \times \qty{700(15)}{kHz}$) for the low-density case (filled lighgreen area). This confirms that operating at $(f, d) = (\qty{850}{kHz}, 65\%)$ maintains an effective pulse area near $\pi$, successfully approaching the predicted selection bandwidth despite the higher repetition rate. Furthermore, it demonstrates that $s_\text{sw}$ can be tuned to maintain consistent performance across different density regimes, proving that the SVS technique remains highly effective even at high atomic densities.

\section{Experimental implementation}

We experimentally implement the SVS scheme using a hot atomic beam of strontium (${}^{88}\text{Sr}$) and empirically verify the numerical predictions. The atomic source features an oven with a nozzle composed of an array of collimation tubes that results in partial collimation of the atomic beam. Due to the rectangular geometry of the nozzle array, the emerging beam has a highly anisotropic transverse profile. This geometry is advantageous for maximizing spatial overlap with the narrow cavity mode located further downstream (see Fig.~\ref{Fig:1} b), by aligning the long axis of the array with the cavity axis. 
\\
At the same time, the shape of the atomic beam presents a challenge: along the long axis of the rectangular nozzle, absorption on the broad ${}^1S_0 \rightarrow {}^1P_1$ transition used for conventional red-detuned molasses becomes significant. As a result, conventional molasses cooling is limited in its ability to prepare a large number of atoms with a sufficiently narrow velocity distribution along the cavity mode. This constraint has been a limiting factor in previous experiments that operated under similar conditions~\cite{Fama24}.
\\
In the following, we briefly describe the experimental setup and implementation of the SVS scheme. We then present measurements of the atom number and Doppler broadening of the atomic ensemble extracted from cavity normal-mode spectroscopy and compare these results with those of our SVS simulations. Finally, we provide an experimental performance comparison against conventional molasses cooling.

\subsection{Setup}

The experiment is performed using a thermal beam of strontium atoms emitted from an oven through a rectangular nozzle ($12\,\mathrm{mm} \times 2\,\mathrm{mm}$). The atomic beam propagates vertically along the $x$ axis, against gravity, with a mean velocity determined by the temperature of the oven. A magnetic field $\mathbf{B} = B\,\mathbf{e}_x$, with $B \approx \SI{2}{G}$, defines the quantization axis. For the measurements presented here, the oven is operated at $525^\circ\mathrm{C}$, corresponding to a mean longitudinal velocity of $\bar v_x \approx 345\,\mathrm{m/s}$, see App.~\ref{App:Homogeneous broadening inside the cavity mode}. 
\\
The divergence along the $z$ direction, corresponding to the long axis of the rectangular nozzle and the cavity axis, is defined by the aspect ratio of the microtube array inserted into the nozzle. Each tube has a length of $8\,\mathrm{mm}$, an inner diameter of $\qty{305}{\micro m}$, and an outer diameter of $\qty{414}{\micro m}$. This array provides first-order geometric collimation of the atomic beam~\cite{Fama24,Fama:25,BeliSilva2025CavityAtom}. The cavity mode is located approximately $15\,\mathrm{cm}$ downstream of the nozzle.
\\
The primary objective is to minimize the velocity spread along the cavity axis while maintaining a large number of atoms traversing the cavity mode. To this end, the SVS stage is implemented $20,\mathrm{mm}$ downstream of the oven nozzle. Along the $y$-direction, conventional molasses cooling is applied. In this direction the optical density is low enough for molasses cooling to be effective. The orthogonal molasses is divided into two regions: one below the SVS stage, and one overlapping with the SVS stage. The former beams are constantly on, while the latter are switched on and off synchronously to the sweep beam so that they do not disturb velocity selection. All beam parameters are listed in App. Tab.~\ref{tab:beam_params}.
\\
The SVS beams extend over a vertical interaction length of $L=\SI{2.5}{cm}$ and follow the geometry described in the theory section (see Fig.~\ref{Fig:3}). To achieve a homogeneous intensity profile along the atomic beam propagation direction, a Powell lens combined with a convex lens is used to generate a top-hat beam profile in the atomic beam direction. 
For beams propagating along the $y$ axis, a Gaussian beam profile is sufficient.
Spatial homogeneity is particularly important for the selection beam, since its power-broadened linewidth determines the achievable Doppler compression, see Eq.~\ref{Eq:lin_broad}. 
\\
In contrast to the ideal implementation discussed in the theory section (see Sec.~\ref{sec:velocity-swept electron shelving}), the shelving beam is operated continuously in the experiment and is not temporally gated. This implementation is simpler experimentally and was found to perform reliably in practice. Possible higher-order effects arising from the continuous coupling are not considered in the present model, as the experimental observations remain in good qualitative agreement with the theoretical predictions.
\\
The peak saturation parameters are approximately $s_\text{sw} \approx 5$ for the sweeping beam and $s_\text{sel}\,\sim\,10^4$ for the selection beam, corresponding to an effective power-broadened linewidth of $\Gamma_{\mathrm\mathrm{eff}}/(2\pi)\allowbreak\,\approx\,\SI{750}{\kilo\hertz}$ for the selection beam. This parameter regime provides efficient state transfer while maintaining good velocity selectivity.
\\
Because the sweeping beam is applied only from one side, it can only decelerate half of our initial velocity distribution towards zero velocity along the cavity axis. To increase the fraction of atoms transferred into the shelved state, a push beam, addressing the same transition as the sweeping beam, but propagating opposite to it, is applied in between the oven nozzle and the SVS region (App.Tab.~\ref{tab:beam_params}). This shifts atoms with a transverse velocity of $v_\mathrm{trans} \approx +5$\,m/s, which would otherwise fall outside the capture range of the sweeping beam, to $v_\mathrm{trans} < -15$\,m/s, reversing their $v_z$ velocity component such that they now propagate towards the sweeping beam. This increases the fraction of atoms that can be effectively transferred into the shelved state. We employ this trick to compensate for the fact that our oven nozzle angle is optimized for molasses cooling, sending the atomic beam vertically upwards. An improved experimental design would include an angled oven nozzle, optimized for SVS collimation.
\\
During the SVS process, atoms within the selected velocity class are transferred to metastable ${}^3P_{2,0}$ states. Residual ground-state atoms, either lying outside the capture range or remaining because of imperfect state transfer, can potentially disturb the intended use of the collimated atomic beam. For our applications, this disturbance can be reduced by accelerating these atoms to a sufficiently high velocity along the cavity axis, decoupling them in frequency from the cavity mode. We accelerate these residual atoms with a ``cleaning'' beam directly above the SVS region in the same direction as the sweeping beam. The cleaning beam is resonant with the ${}^1S_0 \rightarrow {}^1P_1$ transition and therefore leaves the shelved population unaffected. It has a waist of approximately $\SI{3.75}{mm}$ along the propagation direction of the atomic beam and a saturation parameter of $s\simeq46$, which leads to a maximum velocity change of $\Delta v_\text{max}> 20$\,m/s ($\Delta \gamma_\text{d}> 30$\,MHz).
\\
About $\SI{1.5}{cm}$ before they reach the cavity mode, shelved atoms can be repumped back to the ground state $\ket{g}=\ket{{}^1S_0}$ to allow Normal-mode-splitting measurement on the $\ket{{}^1S_0} \rightarrow \ket{{}^3P_1}$ transition. This is done using two lasers that address the ${}^3P_{2,0} \rightarrow {}^3S_1$ transitions. Subsequent decay through the ${}^3P_1$ manifold returns atoms to the ground state, from where they interact with the near-resonant cavity TEM$_{00}$ mode. 
\\
Given the finite lifetime of the ${}^3P_1$ state and the limited propagation distance of roughly $x_d=\SI{1.5}{cm}$ before reaching the cavity, the probability of spontaneous decay into the ground state as the atoms travel toward the cavity mode is $1-e^{-\gamma\tau}$ with $\tau = x_d/\bar v_x$, producing a repumping efficiency of approximately $70\%$ under our experimental conditions.

\subsection{Results\label{sec:ExpResults}}

\begin{figure}[t] 
\centering 
\includegraphics[width=1\linewidth]{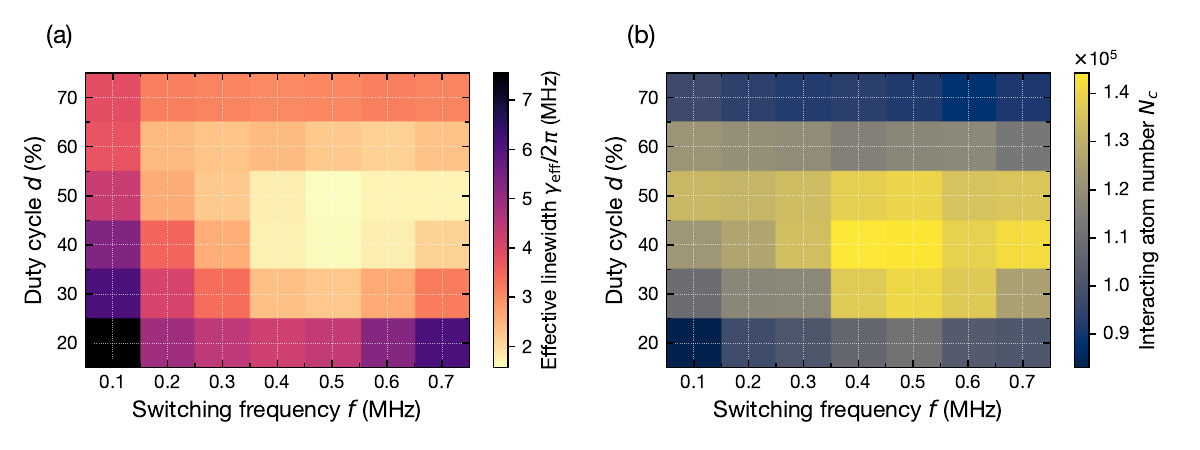} 
\caption{Experimental optimization of duty cycle and switching frequency for swept velocity shelving. The duty cycle $d$ and switching frequency $f$ of the selection- and sweeping beams are varied to find the optimum parameters for large atom number in a narrow longitudinal velocity class. For each parameter set, the effective atomic linewidth $\gamma_\mathrm{eff}$ (a) and the effective number of interacting atoms $N_C$ (b) are extracted from normal-mode splitting measurements on the ${}^1S_0 \leftrightarrow {}^3P_1$ ($m_J=0$) transition. All points are averaged over four independent experimental realizations where each one consist of 128 measurements. A minimum effective linewidth of $2\pi \times(1.57 \pm 0.05)\,\si{\mega\hertz}$ is observed near a duty cycle of $d=50\%$ and a switching frequency of $f=\SI{0.5}{\mega\hertz}$, while the maximum effective atom number, $N_C = (1.40 \pm 0.02)\times10^{5}$, is obtained for a duty cycle of $d=40\%$ and a switching frequency of $f=\SI{0.4}{\mega\hertz}$.}
\label{fig:Pump_push} 
\end{figure}

\subsubsection{Characterization via normal-mode spectroscopy}

To quantify the effectiveness of SVS preparation and compare it with conventional molasses cooling under similar conditions, we require quantitative measures of the Doppler and transit-time broadened ${}^1S_0 \rightarrow {}^3P_1$ linewidth as well as of the atomic density in and flux through the cavity mode. We obtain these quantities from cavity transmission spectroscopy in the strong collective coupling regime.
\\
The goal of the experiment is to reach the regime in which the collective coupling strength $G$ becomes comparable to or exceeds the dominant dissipation rates of the system, namely the cavity field decay rate $\kappa$ and the atomic decoherence rate $\gamma_\mathrm{eff}$. In this regime, the atomic transition hybridizes with the cavity field, forming two coupled normal modes.
\\
Experimentally, this hybridization is observed by injecting weak probe light into the cavity and recording the transmission spectrum (App. Fig.~\ref{fig:Avoided_crossing}). When a collectively coupled atomic ensemble is present, the spectrum has a characteristic double-peaked structure known as normal-mode splitting~\cite{Thompson1992ObservationNormalMode,Raizen1989NormalModeSplitting,Zhu1990VacuumRabiSplitting}. By contrast, in the absence of atoms or with all atoms in the shelving states, the cavity exhibits a single Lorentzian resonance (App.~\ref{App:Bare cavity}). 
\\
The splitting of the two peaks is primarily determined by the collective coupling strength $G$ and therefore by the effective atom number $N_C$ (App.~\ref{App:Atom number, density and total flux trough the cavity}). In contrast, the linewidth and detailed spectral shape reflect homogeneous and inhomogeneous broadening mechanisms, including finite transit-time effects and residual Doppler broadening along the cavity axis (App.~\ref{App:Homogeneous broadening inside the cavity mode}, \ref{App:Velocity distribution in cavity direction}).
\\
By fitting the measured transmission spectra with the model described in App.~\ref{App:Transmission model}, we simultaneously extract the effective atomic linewidth and the interacting atom number. Combining $N_C$ with the known mean beam velocity and the cavity mode volume allows us to determine the atomic density and flux through the cavity. The extracted linewidth further provides a direct measure of the residual longitudinal Doppler broadening. Together, these quantities form the basis for evaluating the performance of the preparation scheme.

\subsubsection{Performance of the SVS scheme}

As a first step, we optimize the duty cycle $d = \tau_\text{sel}f$ and the switching frequency $f = 1/(\tau_\text{sel} + \tau_\text{d})$ of the selection and sweeping sequence, as defined in Sec.~\ref{sec:velocity-swept electron shelving}. The goal is to maximize the effective number of atoms $N_C$ while minimizing the effective linewidth $\gamma_\mathrm{eff}$.
\\
For each parameter set, the probe laser is scanned across the atom–cavity resonance over a range of $\SI{40}{\mega\hertz}$ within $\SI{10}{\milli\second}$, and the transmitted spectrum is recorded. Each spectrum represents an average of 128 repetitions, and the full dataset at each point is fitted with the transmission model described in App.~\ref{App:Transmission model}. Throughout these measurements, the cavity is kept in resonance with the bare atomic transition ($\omega_c = \omega_a$).
\\
The resulting color map is shown in Fig.~\ref{fig:Pump_push}. A global maximum in the collective coupling is obtained at ($d=40\%$, $f=\SI{0.4}{\mega\hertz}$), which yields  $N_C = (1.40 \pm 0.02)\times10^{5}$. The minimum effective linewidth, $\gamma_\mathrm{eff} = 2\pi \times(1.57 \pm 0.05)\,\si{\mega\hertz}$, occurs at similar parameters: ($d=50\%$, $f=\SI{0.5}{\mega\hertz}$). The proximity of these extrema indicates the existence of a global optimum in the parameter space. 

This is consistent with operating near a $\pi$-pulse condition for the selection stage, which optimizes population transfer on-resonance while minimizing off-resonant excitation, ultimately preserving a narrower velocity spread.
\\
To compare with our theoretical simulations (see Fig.~\ref{Fig:4trial}(b)), we note that the theoretical transfer efficiency, $\eta$, is directly related to the measured atom number, $N_\text{C}$. The optimal experimental duty cycle aligns well with the predicted value, whereas the observed optimal switching frequency is lower than expected.
Furthermore, while a distinct global optimum is found within the space of scanned experimental parameters, the theoretical transfer efficiency instead reaches a saturation plateau. Simulations verify that a drop in efficiency eventually occurs at much higher switching frequencies (around \qty{1.5}{MHz}); however, this regime is not reported here as it lies well beyond our experimentally reachable parameters. 
Regarding the final velocity spread, the theoretical value calculated at the simulated optimum ($\gamma_\text{eff} \approx 2\pi \times \qty{0.72}{MHz}$) lies on the same order of magnitude as the experimental minimum ($\gamma_\text{sw} = 2\pi \times \qty{1.22}{MHz}$, see App.~\ref{App:Velocity distribution in cavity direction}), though it is lower by less then a factor of two. 
We attribute this mismatch to two main factors: experimental imperfections omitted from our model and differences in the switching sequence. Primary experimental imperfections include spatial inhomogeneities in optical beams and spurious reflections from imperfect optics. Additionally, the model assumes a simple constant vertical atomic velocity, whereas the experiment involves a velocity distribution.
Given the approximations inherent in this relatively straightforward model, we conclude that the experimental results are in reasonable agreement with the theory and that the technique is adequately captured by the simulation.
\\
Using the relation between $N_C$ and the total number of atoms (App.~\ref{App:Atom number, density and total flux trough the cavity}), we infer that $N \simeq (2.88 \pm 0.06)\times10^{5}$ atoms traverse the cavity mode simultaneously under optimal conditions. This corresponds to an atomic density of $\rho \simeq (7.0 \pm 0.1)\times10^{8}\,\si{cm^{-3}}$ and a mean atom flux of $\Phi \simeq (5.7 \pm 0.1)\times10^{11}\,\si{\per\second}$.

\begin{figure}[t] 
\centering 
\includegraphics[width=\linewidth]{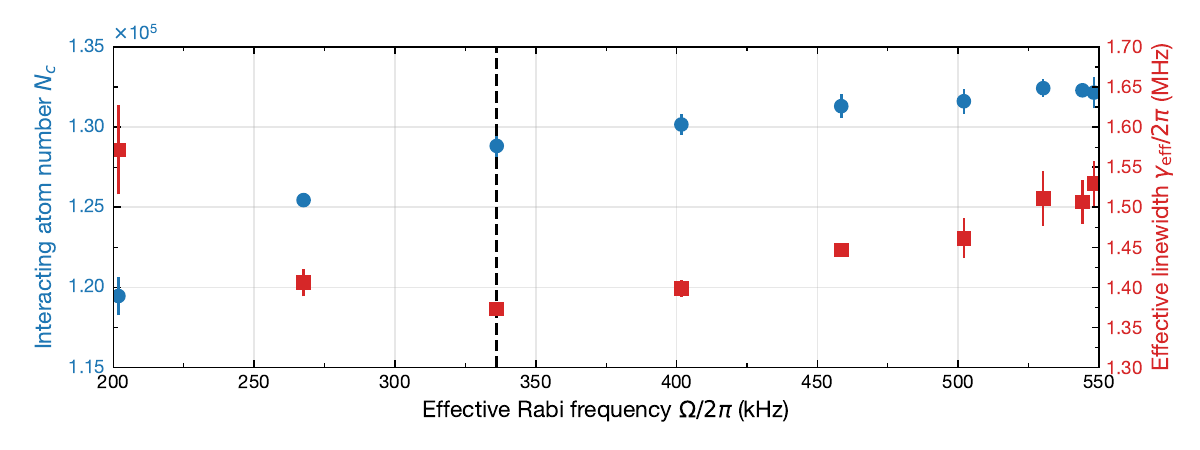} 
\caption{Pump-power dependence of the effective atomic linewidth.
\\
The selection-beam power is varied to reduce the longitudinal velocity spread of the shelved atoms, while keeping the duty cycle fixed at $d=50\%$ and the switching frequency at $f=\SI{0.5}{\mega\hertz}$, corresponding to the optimal parameters identified in Fig.~\ref{fig:Pump_push} for a Rabi frequency of $2\pi \times\Omega=\SI{335}{\kilo\hertz}$ (vertical dashed line).
Increasing the selection beam power leads to power broadening of the selection transition, which in turn determines the width of the selected longitudinal velocity class. For each selection power, the effective atomic linewidth $\gamma_\mathrm{eff}$ (red squares) and the effective number of interacting atoms $N_C$ (blue circles) are extracted from normal-mode splitting measurements using the same measurement and analysis procedure as in Fig.~\ref{fig:Pump_push}. The pump power is expressed in terms of the effective Rabi frequency.
\\
A minimum effective linewidth of $2\pi \times(1.37 \pm 0.05)\,\si{\mega\hertz}$ is achieved while the interacting atom number is reduced by less than $15\%$ compared to its maximum value.}
\label{fig:Power_broadening} 
\end{figure}

Next, we investigate the dependence on the power of the selection beam while fixing the duty cycle and switching frequency at $(d=50\%, f=\SI{0.5}{\mega\hertz})$. As shown in Fig.~\ref{fig:Power_broadening}, the minimum achievable linewidth is primarily governed by power broadening of the selection transition (see Eq.~\ref{Eq:lin_broad}). At low Rabi frequencies, the state transfer becomes inefficient ($\Omega \tau_{\mathrm{sel}} \lesssim \pi$), reducing the shelved population and increasing unwanted acceleration by the sweeping beam. At high power, the velocity selectivity approaches the power broadened transition linewidth.
\\
An optimal compromise is found at an effective Rabi frequency of $\Omega \simeq 2\pi \times\SI{335}{\kilo\hertz}$, where the effective linewidth reaches $\gamma_\mathrm{eff} = 2\pi \times(1.37 \pm 0.05)\,\si{\mega\hertz}$, while the number of interacting atoms is reduced by less than $15\%$ compared to its maximum value.
\\
From the extracted effective linewidth we estimate the residual Doppler contribution along the cavity axis. Assuming a Lorentzian Doppler profile, consistent with the spectral profile of the velocity-selective shelving transition, yields a minimal Doppler linewidth of $\gamma_\text{sw} \approx 2\pi \times\SI{1.02}{\mega\hertz}$. Assuming a homogeneous contribution of $\gamma_{\mathrm{homo}} \approx 2\pi \times\SI{0.35}{\mega\hertz}$, see App.~\ref{App:Velocity distribution in cavity direction}, this corresponds to a longitudinal velocity spread of
\[
v_z^{\mathrm{max}} = 0.7 \pm 0.1~\si{\meter/\second}.
\]
This value is consistent with the effective power-broadened linewidth $\gamma_{\mathrm{eff}}/(2\pi) \simeq \SI{0.75}{\mega\hertz}$ of the selection beam, indicating that the achievable velocity compression is predominantly determined by the power-broadened transition width.

\subsubsection{Comparison with conventional molasses cooling}

Having established the performance of the SVS scheme at an oven temperature of $525^\circ$C ($\approx 800$ K), we now compare it with previous measurements obtained in a similar system employing conventional blue-detuned molasses cooling~\cite{Fama24}.
\\
Under comparable thermal beam conditions, molasses cooling yielded an effective collective atom number of $N_C \simeq 4\times10^{5}$. However, the corresponding velocity distribution exhibited a Gaussian Doppler width of approximately $2\pi \times3\,\si{\mega\hertz}$ together with an additional tilt-induced contribution of $ 2\pi \times4\,\si{\mega\hertz}$. The latter originates from residual misalignment uncompensated by an imbalanced molasses, resulting in a velocity distribution that is no longer centered at zero velocity. Such effects become increasingly pronounced at high atomic densities, where absorption and multiple scattering reduce the efficiency of balanced radiative cooling forces~\cite{Fama24}.
\\
In contrast, the SVS scheme produces a velocity profile with a center value and width, programmable by the selection beam frequency and intensity. We realize an effective Doppler width of $\gamma_\text{sw} \simeq 2\pi \times1\,\si{\mega\hertz}$ and no observable tilt contribution. While molasses cooling reduces the thermal velocity spread through dissipative forces, our method uses SVS, which spectrally filters the velocity classes. The residual linewidth is primarily determined by power broadening of the selection transition, resulting in a controlled and symmetric velocity distribution.
\\
The effective atom number achieved in the present implementation, $N_C \simeq 1.4\times10^{5}$, is approximately three times lower than in the molasses case. However, the transverse cooling region employed here is reduced by approximately a factor of two in the cavity direction, which directly limits the maximum flux of selected atoms. The current configuration therefore prioritizes velocity compression and spectral purity over flux optimization.
\\
An important distinction between the two approaches lies in their behavior at high oven temperatures. In conventional molasses cooling, operation closer to the Knudsen diffusion regime degrades performance due to increased absorption and the breakdown of ideal collimation conditions~\cite{Fama24}. Because SVS relies on swept velocity shelving rather than balanced radiation pressure forces, it remains effective in dense beams, where absorption otherwise limits molasses cooling. The average number of scattered photons in the SVS scheme is significantly reduced as atoms that fulfill the velocity requirement are shelved in dark states, further reducing laser power requirements.
\\
The present results therefore suggest that SVS provides a scalable route toward increasing the collectively coupled atom number at higher oven temperatures while maintaining a narrow longitudinal velocity distribution. Provided that the selection process remains predominantly limited by power broadening, increasing the atomic flux should primarily enhance $N_C$ without increasing the residual Doppler width. A systematic investigation of this scaling behavior is left as the subject of future work.

\section{Conclusions and Outlook}

\subsection{Conclusion}

We have investigated the limitations of optical molasses cooling in high-density atomic beams and demonstrated an alternative cooling strategy based on swept velocity shelving. In conventional molasses, the counter-propagating beams are attenuated as they traverse the dense atomic medium, producing an intensity imbalance that results in a broad velocity distribution. This effect becomes increasingly pronounced at higher atomic densities, where absorption and multiple scattering degrade the balance of radiative forces. This cannot be compensated by simply increasing beam saturation, as that inherently raises the minimum achievable Doppler width.
\\
As an alternative, we have introduced and experimentally demonstrated the SVS scheme in which atoms are first decelerated by a near-resonant sweeping beam and subsequently transferred to a metastable reservoir via a power-broadened narrow-line transition. These two stages are applied in an interleaved sequence. Crucially, the achievable velocity spread in this scheme is nearly independent of the atomic density. For low enough switching frequencies, the width of the selected velocity distribution remains determined primarily by the linewidth, and power broadening, of the selection transition. This represents a qualitative advantage over molasses cooling in the high-density regime.
\\
In our experimental implementation, the velocity distribution of the shelved atoms was characterized via normal-mode splitting in an optical cavity, a diagnostic that directly probes the collectively coupled atoms in the cavity mode and provides access to their velocity distribution. The SVS scheme produces a velocity profile centered at zero velocity, with a minimum effective Doppler linewidth demonstrated at $2\pi \times(1.05 \pm 0.1)$\,MHz (FWHM), approaching the power-broadening limit of approximately $2\pi \times0.8$\,MHz set by the selection beam parameters. These experimental findings are in good qualitative agreement with our theoretical simulations, which successfully model the technique's performance despite geometric and operational simplifications.
\\
These findings compare favorably with the best result obtained under conventional molasses cooling in a previous version of our apparatus, which yielded a velocity distribution with a Doppler width of $\approx 2\pi \times3$\,MHz arising from residual force imbalance at high atomic density~\cite{Fama24}. Although the effective collectively coupled atom number achieved in the present implementation, $N_C \approx 1.4 \times 10^5$, is approximately three times lower than in the case of molasses ($N_C \approx 4 \times 10^5$), this reflects the reduced interaction length chosen in the current geometry rather than a fundamental constraint of the method. Re-establishing the interaction region or operating at higher oven temperatures is expected to recover and surpass the number of atoms reached with optical molasses, while preserving the narrow velocity distribution provided by the SVS scheme.
\\
We note that the SVS approach is not specific to strontium. The scheme requires only three ingredients: a broad transition for efficient deceleration, a narrow transition for velocity-selective transfer, and a metastable shelving state with a sufficiently long lifetime. These conditions are met by a range of alkaline-earth and alkaline-earth-like atoms, and maybe even by select molecular species, making SVS a broadly applicable strategy for producing narrow longitudinal velocity distributions in high-brightness beams.

\subsection{Outlook}

The present work demonstrates SVS along a single spatial direction. A natural extension is to achieve narrow velocity distributions in two dimensions. A direct application of SVS along a second orthogonal axis is not straightforward since atoms already shelved in the metastable state cannot interact with the sweeping or selection beams a second time. However, the metastable state itself opens an attractive alternative route.
\\
In the current experimental configuration, a blue molasses stage applied orthogonal to the cavity axis reduces the transverse velocity spread to approximately $1-2$\,m/s (FWHM), as established by numerical simulations. This provides a suitable starting condition for subsequent narrow-line laser cooling acting directly on the metastable population. Crucially, such cooling can be applied simultaneously with the SVS stage and within the same interaction region, as long as the relevant transitions are spectrally isolated. For strontium, one possibility would be to apply an optical molasses to the metastable atoms: the SVS cycle drives the $^1S_0 \to {^3P_1} \to {^3S_1}$ pathway, from which atoms decay into the $^3P_2$ (and $^3P_0$) metastable states, while the proposed orthogonal cooling operates on the $^3P_2 \to {^3D_3}$ transition and is entirely decoupled from the shelving dynamics and the broad ground state atom background. 
\\
For strontium, two transitions from the metastable $^3P_2$ state are of practical interest. The green $^3P_2 \to {^3D_3}$ transition at $\lambda = 496$\,nm has a natural linewidth of $\gamma_\text{green}/(2\pi) = 9.8$\,MHz, corresponding to a capture velocity of $v_\text{cap} \approx \gamma_\text{green}/k \approx 7$\,m/s and a Doppler temperature of $T_\text{sw} \approx \qty{230}{\micro K}$~\cite{Akatsuka_Three_stage_laser_cooling_of_Sr_atoms}. Since the simulated transverse velocity falls well within this capture range, green molasses on the metastable state represents a direct and technically straightforward next step. 
\\
If a lower velocity spread is required, the IR $^3P_2 \to {^3D_3}$ transition at $\lambda \approx \qty{2.92}{\micro m}$, with $\gamma_\text{IR}/(2\pi) = 57$\,kHz and $T_\text{sw} \approx \qty{1.4}{\micro K}$, could be utilized. In its natural linewidth limit, the capture velocity of $v_\text{cap} \approx 20$\,cm/s is insufficient to capture atoms from the current transverse distribution. However, the effective capture range can be extended by approximately one order of magnitude through power broadening or frequency-modulated SWAP cooling~\cite{Steady_state_magneto_optical_Escudero}, making it also feasible for our parameters. Reliable operation in the IR stage would therefore benefit from first improving the transverse pre-cooling, for instance through a green molasses stage. 
\\
Beyond strontium, this 2D extension generalizes to any system satisfying the SVS requirements and additionally possessing a nearly closed cycling transition within the metastable manifold suitable for narrow-line laser cooling. The combination of these properties makes alkaline-earth and alkaline-earth-like atoms particularly well-suited candidates, and the present results provide a concrete experimental path for pursuing high-phase-space-density with narrow velocity spread in 2D. 

\section*{Acknowledgements and Funding information}
The authors would like to thank Jörg Helge Müller for the suggestion to pursue alternative cooling approaches in order to overcome the optical density limitations of our apparatus. This work has received funding from the European Union’s (EU) Horizon 2020 research and innovation program under Grant Agreement No. 820404 (iqClock project) and No. 860579 (MoSaiQc project). It further received funding from the Dutch National Growth Fund (NGF), as part of the Quantum Delta NL programme. SAS has received funding from the Independent Research Fund Denmark for support under Project No. 0131-00023B. SAS and AS have received funding from the European Union’s (EU) Horizon 2020 research and innovation program under the Marie Skłodowska Curie grant agreement No 101109698 and No 101154998.

\paragraph{Author contributions}
SAS and FS proposed the initial experimental design of SVS cooling. FF, CB and SAS constructed the apparatus. Simulations were performed by FF and the presented data was recorded and analyzed by BH and AS with help by RS. The manuscript was prepared by FF, SAS, AS and BH with help by RS and FS. The project was supervised by SAS and FS. Funding was acquired by SAS, AS and FS.

\begin{appendix}
\numberwithin{equation}{section}
\section{Appendix}

\subsection{Experimental beam parameters}
\label{App:Beam parameters}

The beam parameters for all laser beams used in the experiment are summarised in Tab.~\ref{tab:beam_params}. The beams are grouped according to their function: SVS velocity selection, cleaning and state preparation, and orthogonal molasses cooling.

\begin{table}[h!]
    \centering
    \renewcommand{\arraystretch}{1.6}
    \begin{tabular}{llll}
        \hline
        Beam & Transition & Beam dimensions & Intensity \\
        & $\Delta/2\pi$ &      & $s=I/I_0$ \\
        \hline
        \multicolumn{4}{l}{\textit{SVS beams}} \\
        \hline
        \multirow{2}{*}{Sweep$^a$}
            & $^1S_0 \to {}^1P_1$
            & $L_x$ = \SI{25}{\milli\meter}
            & $\SI{170}{\milli\watt\per\square\centi\metre}$ \\
            & $\SI{-7}{\mega\hertz}$
            & $w_y$ = \SI{1.0}{\milli\meter} & $4$ \\
        \hline
        \multirow{2}{*}{Selection$^a$}
            & $^1S_0 \to {}^3P_1\,(m_J{=}0)$
            & $L_x$ = \SI{25}{\milli\meter}
            & $\SI{29}{\milli\watt\per\square\centi\metre}$ \\
            & $\SI{0}{\mega\hertz}$
            & $w_y$ = \SI{1.5}{\milli\meter} & $\sim\!10^4$ \\
        \hline
        \multirow{2}{*}{Shelving$^{a,b}$}
            & $^3P_1\,(m_J{=}0) \to {}^3S_1\,(m_J{=}{-1})$
            & $L_x$ = \SI{25}{\milli\meter}
            & $\SI{9.3}{\milli\watt\per\square\centi\metre}$ \\
            & $\SI{0}{\mega\hertz}$
            & $w_y$ = \SI{1.5}{\milli\meter} & — \\
        \hline
        \multirow{2}{*}{Push$^a$}
            & $^1S_0 \to {}^1P_1$
            & $w_x$ = \SI{2.5}{\milli\meter}
            & $\SI{385}{\milli\watt\per\square\centi\metre}$ \\
            & $\SI{-10}{\mega\hertz}$
            & $w_y$ = \SI{5.0}{\milli\meter} & $9.5$ \\
        \hline
        \multicolumn{4}{l}{\textit{Cleaning and state preparation}} \\
        \hline
        \multirow{2}{*}{Cleaning$^a$}
            & $^1S_0 \to {}^1P_1$
            & $w_x$ = \SI{1.0}{\milli\meter}
            & $\SI{1950}{\milli\watt\per\square\centi\metre}$ \\
            & $\SI{0}{\mega\hertz}$
            & $w_y$ = \SI{3.75}{\milli\meter} & $48$ \\
        \hline
        \multirow{2}{*}{State prep$^{a,b}$}
            & $^3P_2 \to {}^3S_1$ \& $^3P_0 \to {}^3S_1$
            & $L_z$ = \SI{15}{\milli\meter}
            & 63 \& \SI{29}{\milli\watt\per\square\centi\metre}\\
            & \SI{0}{\mega\hertz} \& \SI{0}{\mega\hertz} 
            & $w_x$ = \SI{0.15}{\milli\meter} & — \\
        \hline
        \multicolumn{4}{l}{\textit{Orthogonal molasses}} \\
        \hline
        \multirow{2}{*}{Molasses (bottom)$^a$}
            & $^1S_0 \to {}^1P_1$
            & $w_x$ = \SI{6.7}{\milli\meter}
            & $\SI{140}{\milli\watt\per\square\centi\metre}$ \\
            & $\SI{-4}{\mega\hertz}$
            & $w_z$ = \SI{6.7}{\milli\meter} & $3.4$ \\
        \hline
        \multirow{2}{*}{Molasses (top)$^a$}
            & $^1S_0 \to {}^1P_1$
            & $w_x$ = \SI{8.9}{\milli\meter}
            & $\SI{48}{\milli\watt\per\square\centi\metre}$ \\
            & $\SI{-4}{\mega\hertz}$
            & $w_z$ = \SI{8.9}{\milli\meter} & $1.2$ \\
        \hline
    \end{tabular}
    \caption{Summary of laser beam parameters for the experimental implementation. The intensity $I$ is the peak value at the Gaussian waist center. The saturation parameter is defined as $s = I/I_\mathrm{sat}$, where $I_\mathrm{sat}$ is the saturation intensity of the respective transition. The detuning is defined as $\Delta = \omega_\mathrm{laser} - \omega_\mathrm{atom}$, with negative values indicating red-detuning. For beams shaped by a Powell lens and cylindrical lens, $L$ denotes the flat-top length and $w$ the Gaussian waist along the orthogonal axis. For Gaussian beams, $w_x$ and $w_y$ are the $1/e^2$ intensity waists. The sweep, selection and top molasses beams are time-sequenced (switched) as described in Sec.~\ref{sec:velocity-swept electron shelving}, all remaining beams are continuous. Entries marked --- are not applicable.\\
    $^a$ Linear polarization $\Rightarrow$ $\pi$ light.\\
    $^b$ The transition is not a closed transition; $I_\mathrm{sat}$ is not defined and $s$ is not quoted.}
    \label{tab:beam_params}
\end{table}

\subsection{Transmission model}
\label{App:Transmission model}
\begin{figure}[h!] 
\centering 
\includegraphics[width=1\linewidth]{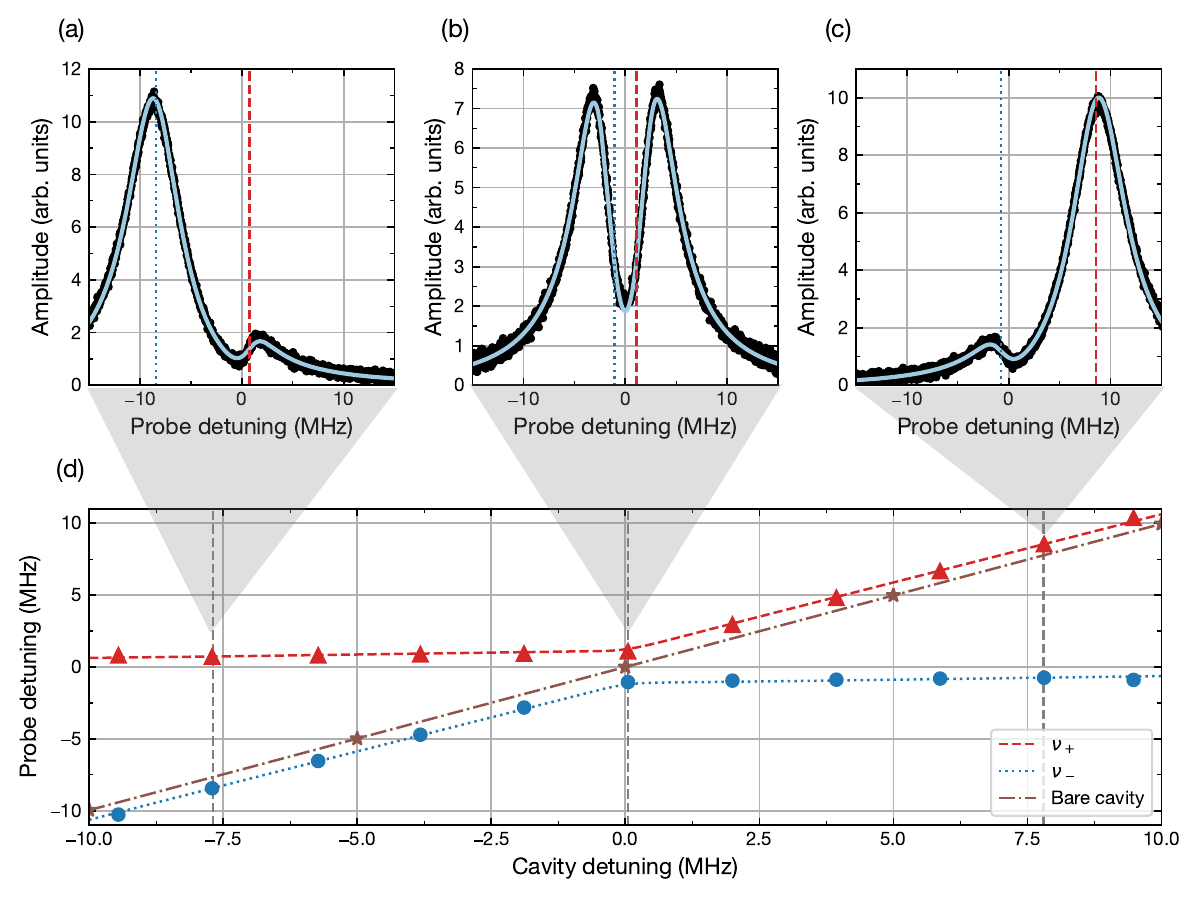} 
\caption{Normal-mode splitting and avoided crossing of the coupled atom--cavity system.
\\
To validate the transmission model, the cavity resonance is detuned from the atomic transition and the cavity transmission is measured by scanning a weak probe laser across the cavity resonance in the strong collective coupling regime, with the atoms prepared in the ground state. Panels (a)-(c) show the measured transmission spectra (black points) for three different cavity detunings; solid gray lines represent fits of the transmission model. The vertical red and blue dashed lines indicate the real parts of the normal-mode eigenfrequencies $\omega_{\pm}$ calculated from the fitted parameters using Eq.~\ref{eq:eigenfreq_lorentzian}.
\\
Panel (d) shows the extracted real eigenfrequencies (red and blue points) as a function of cavity detuning, exhibiting the characteristic avoided crossing due to strong atom--cavity coupling. The dashed lines are calculated using the mean fitted parameters across all detunings and serve as a guide to the eye. For comparison, the bare-cavity response measured in the absence of atoms is shown by the brown stars, demonstrating the expected linear dependence of the cavity resonance on probe detuning (dash-dotted brown line; linear fit to the data).}
\label{fig:Avoided_crossing} 
\end{figure}

Normal-mode splitting of the coupled atom-cavity system is used to extract two key parameters: the effective number of atoms $N_{\mathrm{C}}$ interacting with the cavity TEM$_{00}$ mode at a given time and the effective inhomogeneous broadening along the cavity axis. These quantities provide a direct measurement of the efficiency of the applied laser-cooling techniques and allow us to infer the atomic flux through the cavity and the longitudinal velocity distribution.
\\
To this end, we employ a cavity transmission model based on Refs.~\cite{Thompson1992ObservationNormalMode,Raizen1989NormalModeSplitting,Zhu1990VacuumRabiSplitting}, in which a weak probe laser interrogates the cavity near resonance and the transmitted intensity is detected. In the linear-response regime and under strong collective coupling conditions, defined by
\begin{equation}
G = g_0 \sqrt{N_{\mathrm{C}}} \, \gtrsim \, \kappa ,
\label{eq:collective coupling strength}
\end{equation}
where $G$ is the collective coupling strength, $g_0$ is the single-atom coupling rate, $N_{\mathrm{C}}$ the effective number of atoms coupled to the cavity mode, and $\kappa$ the decay rate of the cavity field, the atomic transition hybridizes with the cavity field and gives rise to two normal modes.
\\
We describe the system using a susceptibility formalism.
The cavity transmission is modeled as
\begin{equation}
T(\omega ) = A \, \left|\chi_c(\omega )\right|^2 + \mathrm{offset},
\end{equation}
where the atom--cavity susceptibility is given by
\begin{equation}
\chi_c(\omega ) =
\frac{1}{(\omega  - \omega_c) + i\kappa/2 - G^2 \chi_a(\omega)},
\label{eq:chic_general}
\end{equation}
and the atomic susceptibility is
\begin{equation}
\chi_a(\omega ) = \frac{1}{(\omega  - \omega_a) + i\gamma_\mathrm{eff}/2}.
\label{eq:chic_atomic}
\end{equation}
Here, $\omega_c$ and $\omega_a$ denote the cavity and atomic resonance frequencies, respectively, and $\gamma_\mathrm{eff}$ is an effective atomic linewidth that accounts for all broadening mechanisms, including Doppler broadening along the cavity axis.
\\
Further, the complex eigenfrequencies of the coupled atom--cavity system correspond to the poles of $\chi_c(\omega)$.
Expressing the frequencies relative to the atomic resonance and allowing for a finite cavity detuning $\Delta\omega _c = \omega_c - \omega_a$, the eigenfrequencies are
\begin{equation}
\omega _\pm =
\frac{1}{2}\left[
\Delta\omega _c - i\frac{\kappa - \gamma_\mathrm{eff}}{2}
\pm
\sqrt{4G^2 + \left(\Delta\omega _c - i\frac{\kappa + \gamma_\mathrm{eff}}{2}\right)^2}
\right].
\label{eq:eigenfreq_lorentzian}
\end{equation}
The real parts of $\omega_\pm$ correspond to the normal-mode eigenfrequencies, while the imaginary parts determine the corresponding linewidths. In the limit $G \gg \kappa, \gamma_\mathrm{eff}$, the normal-mode splitting approaches $\mathrm{Re}(\omega _+ - \omega _-) \simeq 2G$. For overlapping modes, interference between the two pole contributions modifies the observed spectral peaks, and the peak separation no longer directly reflects the collective coupling strength.
\\
To validate the transmission model experimentally, we scan the probe laser frequency across the cavity resonance for several small cavity detunings $\Delta\omega_c$ and record the transmitted intensity, as shown in Fig.~\ref{fig:Avoided_crossing}(a--c).
The spectra are fitted using the transmission model with $G$, $\gamma_\mathrm{eff}$, and $\omega_c$ as free parameters, while the cavity linewidth $\kappa$ is fixed independently from bare-cavity measurements performed in the absence of atoms.
We observe excellent agreement between the model and the experimental data for all detunings.
\\
Using the extracted fit parameters, Eq.~\ref{eq:eigenfreq_lorentzian} yields the expected avoided crossing of the normal-mode frequencies, shown in Fig.~\ref{fig:Avoided_crossing}(d).
For all measurements used to extract atomic fluxes and velocity distributions in the main text, the cavity is tuned on resonance with the atomic transition ($\Delta\omega _c=0$), so that the effective linewidth $\gamma_\mathrm{eff}$ can be used to extract the transverse velocity distribution of the atomic beam (longitudinal in cavity direction) and the collective coupling strength $G$ determines the effective number of atoms traversing the cavity mode.

\subsection{Bare cavity}
\label{App:Bare cavity}
\begin{table}[t]
    \centering
    \renewcommand{\arraystretch}{1.6}
    \begin{tabular}{lll}
        \hline
        Parameter & Symbol / Equation & Value \\
        \hline
        Length & $L$ & $\SI{26.65}{\milli\meter}$ \\
        Waist & $w_{0}$ & $\SI{86.8}{\micro\meter}$ \\
        Linewidth (FWHM) & $\kappa/2\pi$ & $\num{6.79(2)}\,\si{\mega\hertz}$ \\
        Free spectral range & $\Delta\nu_{\mathrm{FSR}} = c/2L$ & $\SI{5.62}{\giga\hertz}$ \\
        Finesse & $\mathcal{F} = \Delta\nu_{\mathrm{FSR}}/\kappa/2\pi$ & $\num{828(2)}$ \\
        Single-atom coupling & $g_0/{2\pi}= -\frac{\hat{\varepsilon}\cdot\mathbf{d}}{h}\sqrt{\frac{\omega\hbar}{2\epsilon_0 V}}$ & $\SI{11.3}{\kilo\hertz}$ \\
        \hline
    \end{tabular}
    \caption{Summary of the cavity parameters. Values without quoted uncertainties are estimated from the cavity design. Note, we define $V=\pi w_0L/4$ as the effective mode volume to account for the gaussian intensity and the standing-wave profile of the TEM$_{00}$ mode~\cite{Single-atom_detection_Poldy_2008}}
    \label{tab:cavity_params}
\end{table}

The bare-cavity response is measured by scanning the probe laser frequency across the empty cavity mode, in the absence of ground state atoms. In this situation, the cavity transmission is well described by a Lorentzian lineshape,
\begin{equation}
T(\omega) = A \,
\frac{1}{1 + 4(\omega  - \omega _c)^2/\kappa^2}
+ \mathrm{offset},
\end{equation}
where $\omega _c$ denotes the cavity resonance frequency and $\kappa$ is the cavity linewidth defined as the full width at half maximum (FWHM).
\\
This expression is consistent with the transmission model introduced in Eq.~\ref{eq:chic_general}. In the absence of atom--cavity coupling ($G=0$), the intensity transmission $|\chi_c(\omega )|^2$ reduces to a purely Lorentzian profile.
\\
The bare-cavity measurement therefore provides a direct and independent determination of the cavity linewidth $\kappa$. In all subsequent fits to spectra exhibiting atom--cavity interaction, $\kappa$ is fixed to this independently measured value and is not treated as a free fit parameter.
\\
In addition, the atomic resonance frequency $\omega_a$ is known from an independent frequency-modulation spectroscopy measurement on the $^1S_0 \leftrightarrow {}^3P_1$ transition, with an uncertainty of roughly $\pm \SI{20}{\kilo\hertz}$. The probe laser is stabilized to this reference, allowing the atomic detuning $\Delta\omega _a = \omega  - \omega _a$ and cavity detuning $\Delta\omega _c = \omega _c - \omega _a$ to be precisely defined in the transmission model.
\\
Finally, by repeating the bare-cavity measurement for different cavity lengths, which are tuned using a piezoelectric actuator, the cavity resonance frequency $\omega _c$ is determined for each experimental configuration.
These values, expressed as $\Delta\omega _c$, are shown as brown points in Fig.~\ref{fig:Avoided_crossing} and are subsequently used as fixed parameters in the analysis of the normal-mode spectra in the main text.

\subsection{Homogeneous broadening inside the cavity mode}
\label{App:Homogeneous broadening inside the cavity mode}
In order to extract the transverse velocity distribution from the measured transmission spectra, it is first necessary to quantify all homogeneous broadening mechanisms that contribute to the effective atomic linewidth. These contributions set a baseline linewidth that must be subtracted before Doppler broadening can be attributed to atomic motion along the cavity axis. This procedure is justified within the present model because both the homogeneous broadening mechanisms and the velocity-selected Doppler broadening are approximated by Lorentzian profiles, such that their convolution results in a Lorentzian lineshape with additive linewidths.
\\
Homogeneous broadening is described by the atomic dephasing rate $1/T_2$, which includes spontaneous emission at rate $\gamma$, stray-light-induced decoherence at rate $\gamma_B$, and finite interaction times with the cavity mode.
Atoms traverse the cavity mode ballistically and interact with the intracavity field only for a finite time, giving rise to transit-time broadening, which constitutes a homogeneous dephasing mechanism.
\\
To estimate the transit time as a function of the oven temperature $T$, we model the atomic beam as operating well below the Knudsen diffusion regime, see Ref.~\cite{Fama24}.
The transverse velocity distribution along the $x$ direction is given by
\begin{equation}
f_x(v_x) = \beta m v_x \exp\!\left(-\beta \frac{m v_x^2}{2}\right),
\label{eq:vxdistr}
\end{equation}
where $\beta = 1/(k_B T)$ and $m$ is the mass of the $^{88}$Sr isotope. Note, we only consider the positive part of the distribution.
The mean transverse velocity is then
\begin{align}
\bar v_x
= \int_0^\infty v_x f_x(v_x)\,\mathrm{d}v_x
= \sqrt{\frac{\pi}{2m\beta}}
= \sqrt{\frac{\pi k_B T}{2m}}.
\end{align}
The mean interaction time with the cavity TEM$_{00}$ mode of waist $w$ is therefore 
\begin{align}
\tau = \frac{2w}{\bar v_x},
\label{eq:tau}
\end{align}
which leads to a homogeneous transit-time broadening
\begin{equation}
2\pi\times\gamma_{\mathrm{T}} \approx \frac{1}{ \tau} =  \left(\frac{ k_B T }{32\pi w^2m}\right)^{1/2},
\end{equation}
for an oven temperature of $\SI{525}{^\circ\mathrm{C}}$, the mean transverse velocity is $\bar v_x \approx \SI{345}{\meter\per\second}$, resulting in a transit-time broadening of $\gamma_{\mathrm{T}} \approx 2\pi\times\SI{315}{\kilo\hertz}$. The oven nozzle is operated approximately $\SI{30}{\kelvin}$ hotter than the oven to prevent clogging of the microtubes. Over the temperature range $\SIrange{500}{560}{^\circ\mathrm{C}}$, the transit-time broadening varies only weakly and remains on the order of $\gamma_{\mathrm{T}} \sim 2\pi\times\SI{300}{\kilo\hertz}$, making it the dominant homogeneous broadening mechanism in the present experiment.
\\
The total homogeneous broadening rate is therefore
\begin{equation}
\gamma_{\mathrm{homo}} = \gamma + \gamma_B + \gamma_{\mathrm{T}}.
\end{equation}
In our setup, this can be approximated as $\gamma_{\mathrm{homo}} \approx 2\pi\times\SI{0.35}{\mega\hertz}$.

\subsection{Velocity distribution in cavity direction}
\label{App:Velocity distribution in cavity direction}
Velocity-selective shelving prepares atoms within a narrow velocity class along the cavity axis, characterized by the Lorentzian transition probability of the optical pumping transition. In this regime, Doppler broadening does not appear as a distinct inhomogeneous contribution but can be absorbed into an effective atomic linewidth $\gamma_\mathrm{eff}$. The atomic susceptibility can therefore be approximated by a Lorentzian, as introduced in Eq.~\ref{eq:chic_atomic}.
The inhomogeneous broadening is then dominated by the residual velocity distribution of the atoms along the cavity axis, leading to Doppler broadening.
For a velocity spread $\Delta v_z$, the corresponding Doppler contribution to the linewidth is
\begin{equation}
\gamma_\text{sw} = k \Delta v_z,
\qquad
k = \frac{2\pi}{\lambda}.
\end{equation}
Fitting the transmission spectra yields the effective linewidth $\gamma_\mathrm{eff}$, from which the Doppler contribution can be isolated by subtracting the homogeneous broadening.
This allows us to estimate an effective maximum longitudinal velocity width
\begin{equation}
v_z^{\mathrm{max}} = \frac{(\gamma_\mathrm{eff} - \gamma_{\mathrm{homo}})\lambda}{2\pi},
\end{equation}
where $\lambda = \SI{689}{\nano\meter}$ is the wavelength of the atomic transition.
\\
This quantity provides a convenient parametrization of the longitudinal velocity spread associated with the selected velocity class and should be interpreted as an effective characterization rather than a thermodynamic temperature.
For the optimal parameters shown in Fig.~\ref{fig:Power_broadening}, we obtain a minimal Doppler linewidth of $\gamma_{\mathrm{sw}} \approx 2\pi\times\SI{1.02}{\mega\hertz}$.
Assuming $\gamma_{\mathrm{homo}} \approx 2\pi\times\SI{0.35}{\mega\hertz}$, this corresponds to a longitudinal velocity spread of
\[
v_z^{\mathrm{max}} = 0.7 \pm 0.1~\si{\meter/\second}.
\]

\subsection{Atom number, density and total flux trough the cavity}
\label{App:Atom number, density and total flux trough the cavity}
Finally, normal-mode splitting provides a direct experimental tool to characterize the number of ground-state atoms prepared within a narrow longitudinal velocity class and interacting with the cavity field.
From the extracted collective coupling strength, we derive effective estimates for the atom number, density, and total atomic flux through the cavity mode.
\\
In our setup, the cavity mode waist varies negligibly along the cavity axis, since the Rayleigh range
\begin{equation}
z_R = \frac{\pi w^2}{\lambda} \approx \SI{34.5}{\milli\meter}   
\end{equation}
exceeds the cavity length. An aperture with diameter $d_{\mathrm{ap}} = \SI{17}{\milli\meter}$ is placed just below the cavity to prevent the cavity mirrors from getting coated with strontium. It defines the macroscopic atomic beam geometry and ensures that only atoms propagating through the central region of the cavity volume contribute to the measured signal, while atoms on trajectories far from the cavity axis are blocked.
\\
Under these conditions, longitudinal variations of the cavity mode waist can be neglected, and the atom--cavity coupling is well approximated by the transverse Gaussian mode profile together with the standing-wave structure along the cavity axis.
The collective coupling strength, defined in Eq.~\eqref{eq:collective coupling strength}, introduces an effective atom number $N_C$ that accounts for the spatial overlap between the atomic ensemble and the cavity TEM$_{00}$ mode.
For atoms uniformly distributed along the standing-wave direction, one obtains $N_C \approx N/2$, where $N$ denotes the total number of atoms within the interaction volume.
This allows us to write
\begin{equation}
N \simeq 2N_C = 2\left(\frac{G}{g_0}\right)^2 .
\label{eq:Atom number}
\end{equation}
Using this estimate, we define an effective average atomic density inside the cavity interaction volume as
\begin{equation}
\rho \approx \frac{N}{\pi w^2 d_{\mathrm{ap}}},
\label{eq:atomic beam density}
\end{equation}
where the volume is determined by the cavity mode waist and the aperture diameter.
The mean atomic flux through the cavity can be estimated from the average interaction time $\tau = 2w/\bar v_x$ as
\begin{equation}
\Phi \approx \frac{N}{\tau} = \frac{N \bar v_x}{2w}.
\label{eq:atomic flux}
\end{equation}
For the optimal parameters shown in Fig.~\ref{fig:Pump_push}, we find that $N \simeq (2.88 \pm 0.06)\times10^5$ atoms are present inside the cavity simultaneously, for a duty cycle of $d=40\%$ and a switching frequency of $f=\SI{0.4}{\mega\hertz}$.
This corresponds to an effective atomic density of $\rho \simeq (7.0 \pm 0.1)\times10^{8}\,\si{cm^{-3}}$ and a mean atomic flux of $\Phi \simeq (5.7 \pm 0.1)\times10^{11}\,\si{\per\second}$.
\end{appendix}


\bibliography{bibliography2}


\end{document}